\documentclass[11pt, oneside]{amsart}
\pdfoutput=1

\usepackage{amsmath,amssymb,amsmath}

\usepackage{color}
\usepackage[table]{xcolor}
\usepackage{dcolumn}
\usepackage{caption}
\usepackage{graphicx}

\usepackage[T1]{fontenc}
\usepackage[utf8]{inputenc}
\usepackage{lmodern}
\usepackage[font=small]{caption}
\usepackage{varwidth}

\usepackage[activate={true,nocompatibility},final,tracking=true,kerning=true,spacing=true,factor=1100,stretch=10,shrink=10]{microtype}
\microtypecontext{spacing=nonfrench}
\usepackage{xspace}

\usepackage{rotating}
\usepackage{caption,subcaption}
\usepackage{float}
\usepackage{psfrag}
\usepackage{tabularx}
\usepackage[hyphens]{url}
\usepackage{wrapfig}
\usepackage{longtable}
\usepackage{verbatim}
\usepackage{array,booktabs,multicol,multirow}

\usepackage[english]{babel}
\usepackage{textcomp}
\usepackage{csquotes}

\usepackage{footnote}
\makesavenoteenv{tabular}
\makesavenoteenv{table}

\usepackage{placeins}

\usepackage{enumitem}
\setlist{leftmargin=7mm}

\usepackage[bookmarks, bookmarksopen, bookmarksnumbered, colorlinks,linkcolor=blue,urlcolor=blue,citecolor=blue]{hyperref}
\usepackage[all]{hypcap}
\definecolor{orange}{rgb}{1.0,0.3,0.0}
\definecolor{violet}{rgb}{0.75,0,1}
\definecolor{darkgreen}{rgb}{0,0.6,0}
\definecolor{cyan}{rgb}{0.2,0.7,0.7}
\definecolor{blueish}{rgb}{0.2,0.2,0.8}
\definecolor{darkblue}{rgb}{0.1,0.1,0.9}
\definecolor{lightgray}{gray}{0.9}

\newcommand{\note}[1]{ {\textcolor{blueish}    { ***Note:      #1 }}}

\usepackage[letterpaper, margin=1in]{geometry}

\sloppypar

\begin{document}

\title[]{Report on the Fourth Workshop on Sustainable Software for Science: Practice and Experiences (WSSSPE4)}

\author{Daniel~S.~Katz$^{(1)}$,
Kyle~E.~Niemeyer$^{(2)}$,
Sandra~Gesing$^{(3)}$,
Lorraine~Hwang$^{(4)}$,
Wolfgang~Bangerth$^{(5)}$,
Simon~Hettrick$^{(6)}$,
Ray Idaszak$^{(7)}$,
Jean~Salac$^{(8)}$,
Neil~Chue~Hong$^{(9)}$,
Santiago~N\'u\~nez-Corrales$^{(10)}$,
Alice~Allen$^{(11)}$,
R.~Stuart~Geiger$^{(12)}$,
Jonah~Miller$^{(13)}$,
Emily~Chen$^{(14)}$,
Anshu~Dubey$^{(15)}$,
Patricia~Lago$^{(16)}$
}

\thanks{{}$^{(1)}$ National Center for Supercomputing Applications (NCSA) \&
Department of Computer Science  \&
Department of Electrical and Computer Engineering  \&
School of Information Sciences (iSchool),
University of Illinois, Urbana--Champaign, IL, USA; d.katz@ieee.org; ORCID: 0000-0001-5934-7525}
\thanks{{}$^{(2)}$ School of Mechanical, Industrial, and Manufacturing Engineering,
Oregon State University, Corvallis, OR, USA; kyle.niemeyer@oregonstate.edu; ORCID: 0000-0003-4425-7097}
\thanks{{}$^{(3)}$ Center for Research Computing \& Department of Computer Science and Engineering,
University of Notre Dame, IN, USA; sandra.gesing@nd.edu; ORCID: 0000-0002-6051-0673}
\thanks{{}$^{(4)}$ University of California, Davis, CA, USA; ljhwang@ucdavis.edu; ORCID: 0000-0002-1021-3101}
\thanks{{}$^{(5)}$ Department of Mathematics, Colorado State
  University, Fort Collins, CO, USA; bangerth@colostate.edu; ORCID: 0000-0003-2311-9402}
\thanks{{}$^{(6)}$ The Software Sustainability Institute, Southampton, UK; s.hettrick@software.ac.uk; ORCID: 0000-0002-6809-5195}
\thanks{{}$^{(7)}$ RENCI, University of North Carolina at Chapel Hill, Chapel Hill, NC, USA; rayi@renci.org; ORCID: 0000-0002-3444-7615}
\thanks{{}$^{(8)}$ University of Virginia, Charlottesville, VA, USA; jeansalac@virginia.edu; ORCID: 0000-0001-5971-9333}
\thanks{{}$^{(9)}$ University of Edinburgh, UK; N.ChueHong@epcc.ed.ac.uk;
ORCID: 0000-0002-8876-7606}
\thanks{{}$^{(10)}$ Illinois Informatics Institute (I3) and National Center for Supercomputing Applications (NCSA), University of Illinois at Urbana-Champaign, IL, USA;
nunezco2@illinois.edu;
ORCID: 0000-0003-4342-6223}
\thanks{{}$^{(11)}$ Astrophysics Source Code Library, Astronomy Department, University of Maryland, College Park, MD, USA; aallen@ascl.net; ORCID: 0000-0003-3477-2845}
\thanks{{}$^{(12)}$ University of California, Berkeley, Berkeley Institute for Data Science (BIDS), Berkeley, CA, USA; stuart@stuartgeiger.com; ORCID: 0000-0001-7215-0532}
\thanks{{}$^{(13)}$ Perimeter Institute for Theoretical Physics \&
Department of Physics, University of Guelph,
Guelph, ON, Canada;
jmiller@perimeterinstitute.ca; ORCID: 0000-0001-6432-7860}
\thanks{{}$^{(14)}$ University of Illinois at Urbana-Champaign, USA; echen35@illinois.edu; ORCID: 0000-0003-2578-0296}
\thanks{{}$^{(15)}$ Mathematics and Computer Science Division,
Argonne National Laboratory,
Lemont, IL, USA \&
University of Chicago,
Chicago, IL, USA;
Adubey@anl.gov;
ORCID: 0000-0003-3299-7426}
\thanks{{}$^{(16)}$ Vrije Universiteit Amsterdam, The Netherlands; p.lago@vu.nl; ORCID: 0000-0002-2234-0845}

\begin{abstract}
This report records and discusses the Fourth Workshop on Sustainable
Software for Science: Practice and Experiences (WSSSPE4). The report
includes a description of the keynote presentation of the workshop, the
mission and vision statements that were drafted at the workshop and
finalized shortly after it, a set of idea papers, position papers, experience
papers, demos, and lightning talks, and a panel discussion.  The main part of the report
covers the set of working groups that formed during the meeting, and for each,
discusses the participants, the objective and goal, and how the objective can
be reached, along with contact information for readers who may want to join
the group. Finally, we present results from a survey of the workshop attendees.


\end{abstract}

\maketitle
\newpage

\section{Introduction} \label{sec:intro}

The Fourth Workshop on Sustainable Software for Science: Practice and Experiences
(WSSSPE4)\footnote{\url{http://wssspe.researchcomputing.org.uk/wssspe4/}} was
held over 2 1/2 days on 12--14 September 2016 in Manchester, England.
This location and date was selected so that WSSSPE4 immediately preceded the
First Research Software Engineers (RSE) Conference.
Previous events in the WSSSPE series include
WSSSPE1\footnote{\url{http://wssspe.researchcomputing.org.uk/wssspe1/}}~\cite{WSSSPE1-pre-report,WSSSPE1},
held in conjunction with the SC13 conference;
WSSSPE1.1\footnote{\url{http://wssspe.researchcomputing.org.uk/wssspe1-1/}}, a
focused workshop organized jointly with the SciPy
conference\footnote{\url{https://conference.scipy.org/scipy2014/participate/wssspe/}};
WSSSPE2\footnote{\url{http://wssspe.researchcomputing.org.uk/wssspe2/}}~\cite{WSSSPE2-pre-report,WSSSPE2},
held in conjunction with SC14;
WSSSPE2.1\footnote{\url{http://wssspe.researchcomputing.org.uk/wssspe2-1/}}, a
focused workshop organized again jointly with
SciPy\footnote{\url{http://scipy2015.scipy.org/ehome/115969/286469/}};
and WSSSPE3\footnote{\url{http://wssspe.researchcomputing.org.uk/wssspe3/}}~\cite{WSSSPE3},
held in Boulder, Colorado, USA.
Note that the first WSSSPE workshop was named
``Working towards
Sustainable Software for Science: Practice and Experiences,'' which remains the meaning
of the WSSSPE group, but the workshops after that were named
``Workshop on Sustainable
Software for Science: Practice and Experiences.'' Together these reflect
that WSSSPE is both a community and a set of workshops.


The WSSSPE4 workshop included multiple mechanisms for participation and
encouraged team building around solutions. WSSSPE4 strongly encouraged participation
of early-career scientists, postdoctoral researchers, graduate students,
early-stage researchers, and those from underrepresented groups,
with funds provided to the conference organizers by the National Science
Foundation (NSF), the Gordon and Betty Moore Foundation, the Alfred P.~Sloan Foundation, and the Software
Sustainability Institute (SSI) to support the travel of potential participants
who would not otherwise be able to attend the workshop. These
funds allowed 29 additional people to attend and participate. A subset of the
organizing committee reviewed 44 applications for travel support and
competitively selected the awardees, including 11 students and five early-career
researchers. In addition, the travel award subcommittee tried to increase the
diversity of applicants by forwarding the notice of travel support to
organizations including PyLadies\footnote{\url{http://www.pyladies.com}},
Django Girls\footnote{\url{https://djangogirls.org}},
Women Who Code\footnote{\url{https://www.womenwhocode.com}}, and
Women in HPC\footnote{\url{http://www.womeninhpc.org}}.

WSSSPE4 also included two professional event organizers/facilitators from
KnowInnovation who helped the organizing committee plan the workshop agenda,
and during the workshop, they actively engaged participants with various
tools, activities, and reminders.

At the workshop, we also introduced a Code of Conduct (CoC).%
\footnote{\url{http://wssspe.researchcomputing.org.uk/wssspe4/code-of-conduct/}}
The CoC was conceived for the workshop itself; however, we intend it
to also be a guideline for the community of scientists that WSSSPE
supports, and their personal and online interactions (e.g., on
Twitter, in email lists, in the Slack team). The WSSSPE4 CoC is based on the
FORCE11 conference CoC~\cite{FORCE11:CoC}, in turn based on the Code4Lib
CoC~\cite{Code4Lib:CoC}.
The main guidelines of the CoC are:
\begin{quote}
    WSSSPE events are community events intended for networking and collaboration
    as well as learning. We value the participation of every member of the
    community and want all attendees to have an enjoyable and fulfilling
    experience. Accordingly, all attendees are expected to show respect and
    courtesy to other attendees throughout the event and in interactions online
    associated with the event.

    The WSSSPE event organizers are dedicated to providing a harassment-free
    experience for everyone, regardless of gender, gender identity and
    expression, age, sexual orientation, disability, physical appearance,
    body size, race, ethnicity, religion (or lack thereof), technology choices,
    or other group status.

    To make clear what is expected, everyone taking part in WSSSPE events and
    discussions---speakers, helpers, organizers, and participants---is required
    to conform to the following Code of Conduct.

    \begin{itemize}
    \item Communicate appropriately for a professional audience including
    people of many different backgrounds. Sexual language and imagery are not
    appropriate for any event.

    \item Be kind to others. Do not insult or put down other attendees. Be
    mindful of jargon, which can sometimes exclude others from engaging in the
    discussion.

    \item Behave professionally. Remember that harassment and sexist, racist,
    ageist, or exclusionary behavior are not appropriate.
    \end{itemize}
\end{quote}

The CoC was introduced at the beginning of WSSSPE4, with the CoC subcommittee
and a general email address introduced for reporting concerns or incidents, or
asking questions.  There was one concern mentioned to the CoC subcommittee
after the first half day,
regarding how part of the workshop was being run, and we changed the workshop
to address this.

This report is based on the events at the workshop and the submitted materials.
These events included discussion of the call for papers (\S\ref{sec:preworkshop}), the WSSSPE mission and vision (\S\ref{sec:mission}), a keynote (\S\ref{sec:keynote}),
a set of presentations (\S\ref{sec:papers} and \S\ref{sec:lightning}), a panel (\S\ref{sec:panel}), and a number of working groups (\S\ref{sec:WGs}).  One full day of the
workshop was spent with participants in the working groups, which occurred in parallel
with each other.  And each of the working groups left with a plan for how they could move
forward.
This report also mentions the Slack channel created for further discussions (\S\ref{sec:slack}), and it highlights an attendee survey (\S\ref{sec:survey}) before concluding (\S\ref{sec:conclusions}).
Appendices list the organizers (\S\ref{sec:orgcom}), attendees(\S\ref{sec:attendees}), travel award recipients (\S\ref{sec:awardees}), and program committee (\S\ref{sec:progcom}), and detail the attendee survey (\S\ref{sec:survey_details}).

\section{Call for participation} \label{sec:preworkshop}

WSSSPE4 was based on the work done at WSSSPE1, WSSSPE2, and WSSSPE3, but aimed
at producing working groups that better continued working after the workshop ended.
In addition, based on feedback after WSSSPE3, it became clear that WSSSPE attendees
had two different motivations in participating.  One motivation was to make a better future
for research software, and the other was to immediately do better research software development.
This led to the idea of WSSSPE4 being partially divided into two tracks:

\begin{quote}
    \textbf{Track 1 -- Building a sustainable future for open-use research
    software} has the goals of defining a vision of the future of open-use
    research software, and in the workshop, initiating the activities that are
    needed to get there. The idea of this track is to first think about where
    we want to be 5 to 10 years from now, without being too concerned with
    where we are today, and then to determine how we can move to this future.

    \noindent \textbf{Track 2 -- Practices \& experiences in sustainable scientific software}
    has the goal of improving the quality of today's research software and the
    experiences of its developers by sharing best practices and experiences.
    This track is focused on the current state of scientific software and what
    we can do to improve it in the short term, starting with where we are today.
\end{quote}

The call for participation requested
idea papers, which present implementable proposals related to Track~1;
position papers, which are longer, not previously published papers related to
Track~2 specifically discussing what we can do to improve sustainable scientific
software in the short term, starting with where we are today;
experience papers, longer papers related to Track~2 that discuss current
practices and experiences and how they have been used to improve the quality of
today's research software and/or the experiences of its developers;
demos, brief papers describing a tool or
process that would be demonstrated, relevant to Track~2 that improves the quality of today's research
software and\slash or the experiences of its developers; and
extended abstracts describing lightning talks,
where each author could make a brief statement about work that either had been
done or was needed.

The following list of contribution topics was suggested, though this was not intended to limit potential submissions:

%
\begin{itemize}
\renewcommand{\labelenumi}{\textbf{\theenumi}.}
\setlength{\rightmargin}{1em}

\item Development and Community
\begin{itemize}
    \item Best practices for developing sustainable software
    \item Models for funding specialist expertise in software collaborations
    \item Software tools that aid sustainability
    \item Academia/industry interaction
    \item Refactoring/improving legacy scientific software
    \item Engineering design for sustainable software
    \item Metrics for the success of scientific software
    \item Adaptation of mainstream software practices for scientific software
\end{itemize}

\item Professionalization
\begin{itemize}
    \item Career paths
    \item RSE as a brand
    \item RSE outside of the UK or Europe
    \item Increase incentives in publishing, funding and promotion for better software
\end{itemize}

\item Training
\begin{itemize}
    \item Training for developing sustainable software
    \item Curriculum for software sustainability
\end{itemize}

\item Credit
\begin{itemize}
    \item Making the existing credit and citation ecosystem work better for software
    \item Future credit and citation ecosystem
    \item Software contributions as a part of tenure review
    \item Case studies of receiving credit for software contributions
    \item Awards and recognition that encourage sustainable software
\end{itemize}

\item Software publishing
\begin{itemize}
    \item Journals and alternative venues for publishing software
    \item Review processes for published software
\end{itemize}

\item Software discoverability/reuse
\begin{itemize}
    \item Proposals and case studies
\end{itemize}

\item Reproducibility and testing
\begin{itemize}
    \item Reproducibility in conferences and journals
    \item Best practices for code testing and code review
\end{itemize}

\end{itemize}

Submissions to WSSSPE4 comprised
19 lightning talks,
4 idea papers,
3 position paper,
5 experience papers,
and
3 demos.
Most of the submissions were accepted, since the goal of WSSSPE is always to
take in as many relevant inputs as possible, and to encourage their authors to
participate in sharing and implementing their ideas.
Specifically,
19 lightning talks,
4 idea papers,
3 position papers,
2 experience papers,
and
3 demos
were accepted. (Two of the submitted lightning talks and one submitted experience paper were rejected, while two more submitted experience papers
were accepted as lightning talks.)
The papers are discussed in Section~\ref{sec:papers},
and the lightning talks are discussed in Section~\ref{sec:lightning}.
The papers and lightning talks have been published as a volume in the CEUR Workshop Proceedings~\cite{WSSSPE4-proceedings}.

\section{Mission and vision}\label{sec:mission}


Going into WSSSPE4, the WSSSPE organization had not had a formal mission or vision statement.
The organizers developed a strawhorse, which was presented to the participants early in the meeting.
The presentation included guidelines as well as examples from other similar communities.
The guidelines were that in general, an organization's mission should stand the test of time and state what is wrong and how the organization is going to fix it; its vision should imagine what the world would look like if the organization is successful; and from which, focus areas could be used to establish the scope of the organization along with its goals.

The participants were given time after the presentation to write down their comments.
Seven people volunteered to work on revising the mission and vision statements, incorporating these comments and presenting back to the community.
Based on this feedback, the committee redrafted the mission and vision.
The committee added focus areas to clarify the organization's breadth and to keep the mission and vision simple and long-lasting.
Comments after the close of the meeting were incorporated into the statements.

A final draft was put on GitHub (\url{https://github.com/WSSSPE/mission-vision}) and advertised
via the WSSSPE mailing list, Facebook group, and Twitter.
After 2 weeks without suggested changes, the final statements, as shown below, were added
to the WSSSPE web page (\url{http://wssspe.researchcomputing.org.uk/about-wssspe/}).

{\bf Mission.}
WSSSPE is an international community-driven organization that promotes sustainable research software by addressing challenges related to the full lifecycle of research software through shared learning and community action.

{\bf Vision.}
We envision a world where research software is accessible, robust, sustained, and recognized as a scholarly research product critical to the advancement of knowledge, learning, and discovery.

{\bf Focus areas.}
WSSSPE promotes sustainable research software by positively impacting:
\begin{itemize}
\item {\bf Principles and Best Practices}. Promoting best practices in sustainable software
\item {\bf Careers}. Developing and supporting career paths in research software development and engineering
\item {\bf Learning}. Engaging in activities to promote peer learning and interaction
\item {\bf Credit}. Ensuring recognition of research software as an intellectual contribution equal to other research products
\end{itemize}

\textbf{Definitions:}
\emph{Sustainable software} has the capacity to endure such that it will continue to
be available in the future, on new platforms, meeting new needs.
The \emph{research software lifecycle} includes:
\begin{itemize}
\item acquiring and assembling resources (including funding and people) into teams and communities
\item developing software
\item using software
\item recognizing contributions to and of software
\item maintaining software
\end{itemize}

\section{Keynote}\label{sec:keynote}


The keynote was given by Patricia Lago and entitled ``The legacy of unsustainable software''.
Sustainability is broadly associated with natural ecologic systems. When we translate the notion of sustainability to software solutions, however, we often confuse {\em impact in a certain point in time} with {\em balanced and durable effects}. In addition, software sustainability adds a fourth dimension to environmental, social and economic aspects: technical sustainability, and hence higher complexity~\cite{Lago2015}.

Supporting technical sustainability in (scientific) software is very near to supporting properties like longevity and resilience, i.e., the ability of a software system or application to fulfill its intent~\cite{Huisman2016}, as planned independently from time passing or from changes in its operational context. Technical sustainability has never been easy, but it is even more of a challenge nowadays with the increasing demands on software solutions to support many aspects of our daily lives.

The keynote discussed the challenges (and some related ongoing research) of combining technical and environmental sustainability. This provided a complementary angle to the workshop discussions. It addressed software energy efficiency (i.e., how to develop software that consumes less energy, like mobile apps that should consume less device-battery) and software energy-awareness (i.e., how to develop software that can actively decide between energy efficiency and other goals such as performance).

The keynote used a few myths to elaborate on such challenges and related research. Among them:

\begin{itemize}
\item[] {\bf Myth \#1: Energy-efficient hardware will solve the issue}. At the implementation level, to make software use resources effectively, we need to first understand the impact of different coding practices on the way software exploits computing resources (hence energy to feed them). In addition, when we have to deal with complex software or new projects, we must understand ``the big picture,'' hence we must reason at the design or architecture level. Maybe the software is not implemented yet, or maybe we must identify which design decisions are more or less beneficial for energy consumption. To this aim, we capture best practices in reusable patterns and so-called architectural tactics. These document generic designs, specific implementations, metrics, and measures.\\

\item[] {\bf Myth \#2: Extensive experimentation will provide the needed know how}. While empirical methods to design and execute software engineering experiments are mature and well known, they are insufficient to handle the too many variables and uncertainties for modern software to be sustainable. We need new empirical methods that help us gather evidence faster~\cite{Procaccianti2015}. This would probably entail relaxing the too many constraints that govern current rules for mitigating threats to validity, and (equally importantly) collaboration- and experience-sharing to build a common knowledge base. In addition, we should exploit better (and more truthful) visualizations of experimentation results, to communicate on advances (and less promising directions.)\\

\item[] {\bf Myth \#3: With massive migration of software systems to cloud-based environments, we do not need to invest in a ``greener cloud'' anymore}. The use of cloud-based services promises large-scale optimizations in terms of the resources used by our software applications and related investments~\cite{green-cloud-sw}. Unfortunately, cloud provisioning heavily relies on over-capacity rather than intelligent governance, and so far has invested very little in innovation for sustainability. We need to build smarter software that both manages resources more efficiently in the cloud, better understands the changes in the software operational context (including user requirements), and dynamically adapts to address sustainability intents~\cite{Procaccianti2016-2} (both in the cloud and on customer's premises.)\\

\end{itemize}

\section{Papers and demos} \label{sec:papers}


WSSSPE4 included the presentation of 12 10-minute talks (based on four idea papers, three position papers,
two experience papers, and three demos) that addressed a wide range of topics around
sustainability for software in science. These talks covered three main areas:
\begin{enumerate}
\item The specifics of the academic environment for software development and for so-called
research programmers or research software engineers;
\item The characteristics and needs of research software in general such as the lifecycle of
software in science, preservation, curation, sharing, and reproducibility challenges; and
\item Lessons learned from or visions for use cases of scientific software and communities
working with and\slash or on a software package.
\end{enumerate}

%
%

The following four papers belong to the first area: the academic environment.
They address diverse aspects ranging from incentives for quality software to
advocating a professional society for research software to roles and degrees for
research software engineers. For talks with multiple authors, an
\textsuperscript{\textasteriskcentered} indicates the author who presented the talk.

\begin{itemize}
\item \textbf{Michael Heroux: \emph{Idea paper: Sustainable \& Productive:
Improving Incentives for Quality Software}}~\cite{Heroux:2016ws}.
The author presents that improving quality of scientific software is a measure
to address the increased complexity in both modeling and software. To support
quality, it is essential to provide Computational Science and Engineering teams
with publication venues, funding, and  professional recognition. Broad
improvements in sustainability and productivity can be only achieved when
publishers, funding agencies and employers raise their expectations for
software quality. Additionally, software community leaders are in the unique
position to positively impact software quality by working on establishing
incentives that will spur creative and novel approaches to improve developer
productivity  and software sustainability.

\item \textbf{Gabrielle Allen: \emph{Idea Paper: Establishing a Professional
Society for Research Software}}~\cite{GAllen:2016ws}.
This paper advocates to create a professional society or extend the scope of an
existing organization in research software. The author presents the motivation,
a variety of potential roles of such a society and diverse potential funding
mechanisms to be able to establish a sustainable organization. She gives an
overview on similar activities as well as existing organizations and aims at
stimulating discussion among potential members and stakeholders.

\item \textbf{Olivier Philippe\textsuperscript{\textasteriskcentered}, Simon Hettrick, and Neil Chue Hong:
\emph{Experience Paper: Preliminary analysis of a survey of UK Research
Software Engineers}}~\cite{Philippe:2016ws}.
This paper presents results from a survey conducted on the new role of research
software engineers in academia. The survey provides demographic information
about the education, field, gender, job satisfaction, and career plans of the people
in this community. The outcome is that members of the community  were found to be
highly educated, derive mainly from the hard sciences, and are predominantly male.
Respondents reported satisfaction in their jobs, but indicated that career progression
is both difficult and opaque. The authors urge a continued discussion about the experience
of research software engineers and recommend further investigation into this
important community.

\item \textbf{Christopher Gwilliams: \emph{Demo: Using industrial engagement to
create and develop research ties within academia}}~\cite{Gwilliams:2016ws}.
This demo focuses on the benefits of including a  new  software engineering
degree that is influenced by, and works directly with,
industry.  Cardiff  University  introduced  a  new  industry-focused  degree,
Bachelor of Science in Applied Software Engineering (ASE) in September
2015. The engagement with industry allows students to learn through live
projects  with  real-world  applications and forge links for career opportunities.
Also, industry benefits from this course that provides an entry point for industry
to engage with the university and to develop a network of contacts between industry
and academia, which is also a benefit for the university and allows to identify
potential opportunities for new research projects and collaborators. \\

\end{itemize}

The second main area, concerned with characteristics and needs of research software, includes the following five papers:

\begin{itemize}

\item \textbf{Anshu Dubey\textsuperscript{\textasteriskcentered} and Katherine Riley: \emph{Experience Paper: Software
Engineering and Community Codes Track in ATPESC}}~\cite{Dubey1:2016ws}.
The Argonne Training Program in Extreme Scale
Computing (ATPESC) was started by Argonne National Laboratory with the objective
of expanding the ranks of better prepared users of high-performance computing (HPC) machines.
One of the unique aspects of the program was inclusion of a
software engineering and community codes track. The inclusion
was motivated by the observation that the projects with a good
software process were better able to meet their scientific goals.
The paper presents the experience in running the software
track, the motivation, reception, and evolution of the track over the years.
The authors recommend the inclusion of similar tracks in other HPC-oriented training programs.

\item \textbf{Neil Chue Hong: \emph{Position Paper: Why do we need to compare
research software, and how should we do it?}}~\cite{ChueHong:2016ws}.
This position paper sets out the reasons for why different stakeholders,
from users to developers to funders, might wish to undertake the difficult task
to compare research software. The author describes a proposed framework for doing
so (based around measures of accessibility, usability, maintainability, and
portability) which takes into account the possibility of variation between
different communities about how they prioritize different aspects of research software.

\item \textbf{Anshu Dubey\textsuperscript{\textasteriskcentered} and Lois Curfman McInnes: \emph{Idea Paper: The Lifecycle
of Software for Scientific Simulation}}~\cite{Dubey2:2016ws}.
The authors discuss the specifics of the lifecycle of scientific software.
While the software lifecycle is a well-researched topic, one class of projects,
research software, has received relatively
little attention from lifecycle researchers. This
paper formalizes a lifecycle model
for end-to-end scientific application software, featuring loose
coupling between submodels for development of infrastructure
and scientific capability. The authors also invite input from stakeholders
to converge on a model that captures the complexity of this
development processes and provides needed lifecycle guidance to
the scientific software community.

\item \textbf{Francisco Queiroz\textsuperscript{\textasteriskcentered} and Rejane Spitz: \emph{Position Paper:
Collaborative Gamification Design for Scientific Software}}~\cite{Queiroz:2016ws}.
The paper focuses on usability design and presents that gamification, a design
trend for improving scientific software usability, for scientific software should
be facilitated by an open, collaborative design process supported by
conversational media. This approach is compatible with
qualities often attributed to computational science community
regarding openness and collaboration between members of varied
professional backgrounds. They exemplify the use of conversational media for
collaborative design and expect the synergy between collaborators to result in
better usability, greater user acceptance, and adequacy to requirements,
obtaining optimal design solutions in a sustainable way.

\item \textbf{Bruce Childers\textsuperscript{\textasteriskcentered}, Jack Davidson, Wayne Graves, Bernard Rous, and
David Wilkinson: \emph{Position Paper: Active Curation of Artifacts is Changing
the Way Digital Libraries will Operate}}~\cite{Childers:2016ws}.
The paper presents the implications of ``active curation systems'' that provide
executable access to
artifacts and experiments behind published results and enables
their reuse. These systems allow changing and repeating experiments to understand
how an innovation behaves in conditions beyond the ones described in a paper and
address reproducibility challenges. Active curation systems
also  enable accountability and accelerate research progress  by
giving access to complete experimental details. As these systems
take  hold,  it  is  important  to  understand  their  capabilities  and
how digital libraries should be integrated with them. The authors describe a
study underway to explore how best
to do this integration.\\

\end{itemize}

Three contributions focused on the third main area: elucidating lessons learned
from or visions for use cases of scientific software and communities working
with and\slash or on a software package.

\begin{itemize}

\item \textbf{Debashis Ganguly, William C.\ Garrison III, David Wilkinson,
Bruce R.\ Childers, Adam J.\ Lee, and Daniel Mosse\textsuperscript{\textasteriskcentered}: \emph{Demo: Composing,
Reproducing, and Sharing Simulations}}~\cite{Ganguly:2016ws}.
The authors demonstrate their approach of sharing,
reproducing, and composing simulations toward accelerating
research productivity while also improving accountability and
credibility. Specifically, they have developed a case study in
which they compose and share access control simulations in the
form of shareable data store units for cloud systems. They share, compose
and repeat simulations  through a collaborative repository
(OCCAM) and suggest a general simulation framework (SST) that can
accelerate the efforts for the whole community.

\item \textbf{Sameer Shende\textsuperscript{\textasteriskcentered} and Allen Malony: \emph{Demo: Using TAU for
Performance Evaluation of Scientific Software}}~\cite{Shende:2016ws}.
This demo presents the  TAU Performance System for performance evaluation of Scientific
Software written in C++, C, and Fortran. The system addresses performance
technology problems  at three levels: instrumentation,
measurement, and analysis. The TAU framework supports the
configuration and integration of these layers. The authors advocate that effective
exploration of performance will necessarily
require prudent selection from the range of alternative methods TAU provides
to  assemble meaningful performance experiments.

\item \textbf{Janos Sallai\textsuperscript{\textasteriskcentered}, Christopher Iacovella, Christoph Klein, and Tengyu Ma:
\emph{Idea Paper: Development of a Software Framework for Formalizing Forcefield
Atom-Typing for Molecular Simulation}}~\cite{Sallai:2016ws}.
The authors present an essential and error-prone task in the area of Molecular
Dynamics simulations: the efficient selection of correct parameter values for
specific molecules, especially forcefields. They present a framework that
aims to solve this data management issue, proposing a common format for
forcefields that is self-documenting with machine readable,
declarative usage rules. They investigate processes and tools
that are commonly used today in software development (e.g.,
unit testing, verification and validation, continuous integration,
and version control) and reason that they are, with proper infrastructure
support, applicable to forcefield development, as well. The
paper describes how such an infrastructure can tackle managing and evolving
forcefields by the molecular dynamics community, and
proposes a way to encourage and incentivize involvement
by the stakeholders.

\end{itemize}

\section{Lightning talks} \label{sec:lightning}

There were 19 lightning talks presented at WSSSPE4, as follows (in order of
presentation).
For talks with multiple authors, an \textsuperscript{\textasteriskcentered} again indicates the author who presented the talk.
\begin{itemize}[itemsep=1ex]
    \item \textbf{Shoaib Sufi: \emph{The Software Sustainability Institute Fellowship
    programme; supporting the social side of research software}}~\cite{Sufi:2016ws}.
    Sufi, Community Lead at the Software Sustainability Institute (SSI), discussed
    the SSI Fellowship program and the fellows, including their
    demographic breakdown and domains of practice. This program came about
    based on the institute's desire to integrate with the research community---thus
    inspiring change from within---rather than preach to it. Five SSI fellows
    attended WSSSPE4.

    \item \textbf{Sandra Gesing\textsuperscript{\textasteriskcentered}, Maytal Dahan, Linda B.~Hayden, Katherine Lawrence,
    Suresh Marru, Marlon Pierce, Nancy Wilkins-Diehr, and Michael Zentner:
    \emph{The Science Gateways Community Institute - Supporting Communities to
    Achieve Sustainability for Their Science Gateways}}~\cite{Gesing:2016ws}.
    In this talk, Gesing introduced the concept of science gateways: end-to-end software
    solutions for a community that hide underlying complexity. She also posed the need
    for gateways, which include the increased complexity of research and associated
    hardware\slash software, a greater need for reproducibility and openness, and the
    opportunity to integrate research with teaching for better workforce development.
    In fact, gateway users surpassed login-based users in XSEDE systems in 2013,
    and 1.4 times as many users accessed XSEDE via gateways compared with
    login in March 2016. The Science Gateways Community Institute offers support
    related to gateways in five areas: Incubator, Extended Developer Support, Scientific
    Software Collaborative, Community Engagement and Exchange, and Workforce
    Development.

    \item \textbf{Karthik Ram\textsuperscript{\textasteriskcentered}, Noam Ross,
    and Scott Chamberlain: \emph{A model for
    peer review and onboarding research software}}~\cite{Ram:2016ws}.
    This talk described the rOpenSci community's review process for research
    software packages. Submission of a package starts with a pre-submission inquiry
    for fit with the community criteria, then peer reviewers evaluate based on
    usability, quality, and style. Upon acceptance, packages are given an rOpenSci
    README badge and added to the system. Important components include a code of
    conduct, open and non-adversarial review, and that no submission is
    rejected---just out of scope. Remaining challenges include scaling, author
    incentives, and automation of the process.

    \item \textbf{Frank L\"{o}ffler and Steven Brandt\textsuperscript{\textasteriskcentered}:
    \emph{A vision of computing in 10+ years}}~\cite{Loffler:2016ws}.
    This talk began by describing the challenges facing science, and particularly
    computational science, in recent years: the amount and complexity of knowledge
    has increased dramatically, making it difficult for individuals or even teams to
    manage modern, complex problems.
    Brandt submitted an argument for decoupling science simulation and
    the solution of computational problems away from numerical methods and low-level
    optimization. This may be possible with domain-specific languages, or the use
    of modern features in existing languages. This vision also includes reuse of
    basic computing infrastructure, including common data structures, parameter files,
    and high-performance I\slash O. This computing future necessarily requires
    better recognition and reward systems for research software developers (i.e.,
    research software engineers), and ways to make increasingly complex scientific
    papers more understandable.

    \item \textbf{Daniel S.~Katz, Kyle E.~Niemeyer\textsuperscript{\textasteriskcentered},
    Arfon M.~Smith, and the FORCE11 Software Citation Working Group:
    \emph{Software Citation: Process, Principles, and Implementation}}~\cite{Katz:2016ws}.
    This talk described the timeline of activities of the FORCE11 Software Citation
    Working Group and the principles of software citation established by the group.
    Niemeyer, one of the working group co-chairs, also discussed the paper published
    describing the principles~\cite{Smith:2016sc}, and the remaining steps of the
    working group before sunsetting it and spinning up a new Software Citation
    Implementation Working Group.

    \item \textbf{Ray Idaszak\textsuperscript{\textasteriskcentered},
    David G.~Tarboton, Hong Yi, Michael Stealey,
    Pabitra Dash, Alva Couch, Daniel P.~Ames, Jeffery S.~Horsburgh, Tony Castronova,
    Jon Goodall, Mohamed Morsy, Venkatesh Merwade, Mauriel Ramirez, Tian Gan,
    Drew (Zhiyu) Li, Jeff Sadler, Shawn Crawley, Zhaokun Xue, Lan Zhao, Carol Song,
    Christina Bandaragoda: \emph{HydroShare - A Case Study in Software Engineering
    Best Practices and Culture Change for Developing Sustainable Community
    Software}}~\cite{Idaszak:2016ws}.
    HydroShare is a five-year-old online, collaborative system that supports
    open sharing of hydrologic data, analytical tools, and computer models. The
    project involves 10 collaborating institutions that produce tested and reviewed
    community code every two to three weeks, following software engineering best
    practices. Undergraduate and graduate students involved in the project not
    only adopt the best practices exhibited by HydroShare, but see the open,
    collaborative approach to research software development as the norm.
    The talk made recommendations to other community efforts, including the need to
    identify committed, hands-on PIs; evidence-based software engineering for science;
    continuous integration; the importance of professional software development and
    operations practices, even for domain scientists; and to identify technology
    decisions that overlap with career paths.

    \item \textbf{Lorraine Hwang\textsuperscript{\textasteriskcentered},
    Wolfgang Bangerth, Timo Heister, and Louise Kellogg:
    \emph{ASPECT: Hackathons as an Example of Sustaining an Open Source
    Community}}~\cite{Hwang:2016ws}.
    In this talk, Hwang first described the ASPECT (Advance Solver for Problems in
    Earth ConvecTion) open-source, parallel, extensible finite element code
    used to simulate thermal convection. ASPECT began development in 2011 as a
    community-driven project under the Computational Infrastructure for Geodynamics
    (CIG). Community-building activities around ASPECT focused on the traditional
    workshops, tutorials, and webinars, but also on slightly nontraditional,
    multi-day hackathons---which turned out to be one of the most effective
    forms of fostering a community. These start with leaders establishing project
    conventions, demonstrating workflows and relevant technical skills, and
    identifying starter projects. Daily activities involve work plans and
    science\slash coding briefs, with mini-tutorials given as needed. One important
    requirement is that each participant must submit at least one commit. Hwang
    also summarized lessons learned, including that hackathons should last 7--10
    days, include participants with multidisciplinary expertise and at both novice
    and expert levels, have at least one dedicated code reviewer per 10 participants,
    hold daily group updates, organize activities to promote group interactions,
    and enforce free time.

    \item \textbf{Carole Goble: \emph{A Simple Profiling Framework for Software
    User-Producer Reciprocity Review}}~\cite{Goble:2016ws}.
    Goble first described some existing frameworks for measuring the maturity of
    software, including the Software Sustainability Maturity Model, OSS Watch
    Openness Rating, NASA Reuse Readiness Rating, Capability Maturity Model, and
    Qualification and Selection of Open Source software. Then, she proposed a new,
    simpler framework for project reciprocity, drawing on the intentions of users
    and producers of the software, from the viewpoint of the producer.
    Next steps include further definition of the model's characteristics based on
    both theoretical and empirical investigation.

    \item \textbf{Stephan Druskat: \emph{A proposal for the measurement and
    documentation of research software sustainability in interactive metadata
    repositories}}~\cite{Druskat:2016ws}.
    In this talk, Druskat first identified two technical barriers to software
    sustainability: identification of good software, and software discoverability.
    Some of his perspectives were based on experiences in the humanities, a lack
    of funding for computational infrastructure results in little experience with
    sustainability (although this may differ in other domains). Druskat proposed
    metadata repositories as a possible solution, as they can measure and document
    technical sustainability, and document resulting metrics and more general
    features of software. Technical sustainability goals include ensuring software
    existence, preserving potential operation of software, and creating\slash
    retaining possibilities for further development and maintenance of software.
    The next steps of such a system include compiling measurable criteria for
    the sustainability goals (based on qualitative concepts of, e.g.,
    ``sustainability,'' ``maintainability,'' and ``usability''), and
    crowd-sourcing a weighting of these criteria.

    \item \textbf{Mistral Contrastin, Matthew Danish,
    Dominic Orchard\textsuperscript{\textasteriskcentered}, and
    Andrew Rice: \emph{Supporting software sustainability with
    lightweight specifications}}~\cite{Contrastin:2016ws}.
    This talk introduced the tool CamFort for verifying Fortran code based on
    lightweight specifications, which aid in software maintainability and reuse
    through high-level indications of programmer intent.
    Currently, CamFort supports specification (and
    verification) of units and stencil operations. Plans for future specifications
    include automatic test generation based on user-supplied properties, dependency
    specifications, and software contracts to check, e.g., expected value ranges
    and program behavior.

    \item \textbf{Alice Allen, Cecilia Aragon, Christoph Becker, Jeffrey Carver,
    Andrei Chi\c{s}, Benoit Combemale, Mike Croucher, Kevin Crowston, Daniel Garijo,
    Ashish Gehani, Carole Goble\textsuperscript{\textasteriskcentered},
    Robert Haines, Robert Hirschfeld, James Howison,
    Kathryn Huff, Caroline Jay, Daniel S.~Katz, Claude Kirchner, Kateryna Kuksenok,
    Ralf L\"{a}mmel, Oscar Nierstrasz, Matt Turk, Rob van Nieuwpoort, Matthew Vaughn,
    and Jurgen Vinju: \emph{``I solemnly pledge'': A Manifesto for Personal
    Responsibility in the Engineering of Academic Software}}~\cite{AAllen:2016ws}.
    This talk described the activities of the weeklong Dagstuhl Perspectives Workshop
    manifesto on ``Engineering Academic Software,'' which produced a manifesto
    describing a pledge affirming the scholarly value of research software and
    the personal responsibility to act in accordance with that value. The draft
    (personal) manifesto includes 20 pledges, organized into three areas:
    recognition of academic software, academic software development processes,
    and the intellectual content of academic software. The talk concluded with a
    survey for participants to give feedback on the pledges and their importance.

    \item \textbf{Neil Chue Hong\textsuperscript{\textasteriskcentered}:
    \textit{Making it easier to understand research software impact}}~\cite{ChueHong:2016wsb}.
    Chue Hong proposed that the WSSSPE community collaboratively develop a framework
    to better understand (and measure the impact of) research software. This could
    be based on a workflow involving mining the literature for mentions of software,
    identifying linkages between software, and visualizing dependencies between
    software or between software and other research artifacts. Many tools already
    exist in this ecosystem, so Chue Hong proposed that a WSSSPE working group
    collect these tools, summarize use cases, identify missing pieces, and find
    low-hanging fruit to integrate tools or build new ones.

    \item \textbf{Willem Robert van Hage\textsuperscript{\textasteriskcentered},
    Jason Maassen and Rob van Nieuwpoort: \textit{Software Impact Measurement at the
    Netherlands eScience Center}}~\cite{vanHage:2016ws}.
    This talk described eStep (eScience Technology Platform) at the Netherlands
    eScience Center, which includes three components: a software catalogue
    of eScience software, interfaces, libraries,
    tools workflows, applications, etc.; a knowledge base of guides, reports, and
    recommendations for scientific software development; and eScience research
    documents including such as scientific publications, book chapters, and demos.
    The eStep website currently presents automatically-generated information about
    software based on metadata, with a plan for displaying measures of impact and
    quality.

    \item \textbf{Gabrielle Allen, Emily Chen\textsuperscript{\textasteriskcentered},
    Ray Idaszak, and Daniel S.\ Katz: \textit{Report on Software Metrics for
    Research Software}}~\cite{GAllen:2016wsb}.
    In this talk, Chen first summarized some of the outcomes of the WSSSPE3
    Metrics Breakout Group, who discussed that metrics are important for scientific
    impact, discovery, reducing duplication of efforts, and evaluating software.
    The group also summarized barriers to widespread use of metrics: no common
    set of metrics collected, a need for usefulness to be obvious for projects,
    and no common standard for collecting\slash presenting metrics. To find out
    more about practices related to metrics, the authors ran a survey of National
    Science Foundation Software Infrastructure for Sustained Innovation (SI2)
    program investigators, with 19 participants. Important results include the
    observation that most use citations and acknowledgements of software in
    funding proposals, while download numbers are both challenging to collect
    and not useful. Overall, participants generally agreed that metrics are
    important, but said useful metrics are difficult or tedious to collect.

    \item \textbf{Eric L.~Seidel\textsuperscript{\textasteriskcentered}
    and Gabrielle Allen: \textit{Bringing Techniques from Software Engineering
    into Scientific Software}}~\cite{Seidel:2016ws}.
    This talk, given by a senior graduate student in Computer Science\slash
    programming languages research, described several areas of software engineering
    concepts that benefit the development of research software. In particular,
    Seidel focused on software correctness, starting from high-level
    specifications. Correctness can be demonstrated by verifying or proving
    that the software satisfies the specification, testing against a range of
    inputs, or synthesis of code from the specification. However, all of these
    present technical challenges, in addition to the social challenges of
    encouraging collaboration between software engineering and research software
    communities.

    \item \textbf{Iain Emsley\textsuperscript{\textasteriskcentered} and David
    De Roure: \textit{Sustaining the social: Connecting the lives of Drupal
    community}}~\cite{Emsley:2016ws}.
    In this talk, Emsley discussed efforts to develop the community around the
    Drupal content management system, which is supported through the DrupalCon
    series of conferences, user groups, and an association.
    The talk described the use of social coding and software credit to develop
    the Drupal software community, and social documentation to provide
    documentation and tutorials. By analyzing identities of committers, they
    found that the larger community is actually a loosely coupled set of
    heterogeneous communities that use (and sustain themselves) through different
    social mechanisms.

    \item \textbf{Jack Dongarra, Sven Hammarling, Nicholas J.\ Higham, Samuel
    D.\ Relton\textsuperscript{\textasteriskcentered}, Pedro Valero-Lara,
    and Mawussi Zounon: \textit{Creating a Standardised Set of Batched
    BLAS Routines}}~\cite{Dongarra:2016ws}.
    This talk described Batched BLAS (BBLAS), a new standard for computing multiple
    BLAS (Basic Linear Algebra Subproblems) in parallel. Unlike existing parallel
    BLAS algorithms provided by, e.g., Intel MKL, cuBLAS, or MAGMA, BBLAS is
    optimized for smaller matrices, which apply to a variety of problems in areas
    such as machine learning, multifrontal solvers, image processing, fluid
    dynamics, and astrophysics. BBLAS offers nearly a factor of two improvement
    in computing speed compared to non-batched operations. To develop a sustainable
    BBLAS library, the developers have focused on proposing a standard API,
    holding workshops between academics and vendors to reach consensus, and
    performing research into API design decisions on performance.

    \item \textbf{Lukas Breitwieser\textsuperscript{\textasteriskcentered},
    Roman Bauer, Alberto Di Meglio, Leonard Johard,
    Marcus Kaiser, Marco Manca, Manuel Mazzara, Fons Rademakers,
    Max Talanov: \textit{The BioDynaMo Project}}~\cite{Bauer:2016ws}.
    Breitwieser described the BioDynaMo project, a large-scale platform for
    biological simulations that provides access to computational resources while
    hiding the complexity of distributed computing. It also promotes computational
    reproducibility through shared open-access data. The project began as a
    modernization initiative porting an existing simulation software, Cx3D, from
    Java to C++, but also incorporated many modern software engineering practices
    such as hosting and version control on GitHub, automated tests and continuous
    integration via Travis-CI, code coverage reports, and an automated build system.
    In addition, over 1000 students participated in development via the Intel
    Modern Code Developer Challenge competition---the winner achieved a
    performance speedup of 320 times. The project plans to continue improving
    the development process with, e.g., static code checkers and Git commit hooks,
    continued architecture redesign, and verification.

    \item \textbf{Aseel Aldabjan, Robert Haines, and
    Caroline Jay\textsuperscript{\textasteriskcentered}: \textit{How should we
    measure the relationship between code quality and software
    sustainability?}}~\cite{Aldabjan:2016ws}.
    This talk explored the relationship between code quality and the sustainability
    of research software; however, the latter has no concrete definition, thus making
    it challenging to provide guidance on how to achieve sustainability. The authors
    proposed a ``software sustainment'' metric based on the time difference between
    the most recent and initial commits\slash contributions to the codebase.
    Using this metric, they analyzed 3113 Java projects hosted on GitHub, created
    in 2009, and found that 22\% and 35\% had sustainment metrics of 0 and less
    than seven, respectively. Interestingly, the authors also found high
    code quality (measured using a variety of static analytic metrics) across
    all these projects, with little correlation to sustainability.

\end{itemize}

\section{Best practices for scientific software panel discussion} \label{sec:panel}


To investigate best practices for scientific software we brought together a panel of five experts with different perspectives on ``Best Practices for Scientific Software.''
These were: Alice Allen, Editor, Astrophysics Source Code Library, who brought an understanding of the difficulties of organizing a community and curating their software;
Mike Croucher from the University of Sheffield and Rob Haines from the University of Manchester who are both Research Software Engineers with decades of experience of writing code for researchers;
Patricia Lago from Vrije Universiteit Amsterdam, who presented a keynote at the workshop (\S\ref{sec:keynote})  and brought a fresh perspective on software sustainability from the point of view of its impact on society and business; and
Karthik Ram from the Berkeley Institute for Data Science, who brought his perspective on the practicalities of using scientific software to conduct his research as a data scientist.

The panel opened with a question to the floor ``Who in the audience still codes in anger?'' which caused some confusion as, it turns out, that this is a turn of phrase that has not crossed the Atlantic.
After translating this question into the more open, ``Who in the audience still writes code as part of their day-to-day work?'' we were able to confirm the suspicion that the WSSSPE audience is heavily involved writing, not just using, code.
Hence, the panel were going to have to deal with questions from a highly experienced audience.

The panel opened by noting that the world of scientific computing is changing rapidly.
Each panelist was asked for their views on practices which had recently emerged and, if not handled correctly, would become the basis for the next generation's software problems?
This quickly led to a discussion about the dangers of containerization.
These technologies are pushed by some---a minority it is hoped---as a panacea for software sustainability.
The panel and attendees generally agreed that containerization is useful, but should not be seen as the solution to sustainability.
After all, these technologies are just as liable as any to decay.

A later question looked at the common practices from software engineering that the scientific community should be using.
The discussion that followed picked up on the key elements of basic sustainable practice: version control, documentation and ensuring that communities are active and engaged.
When asked whether there was resistance from the research community to adopting these practices, few negative experiences were reported.
From the panel's point of view, it would appear that the problem in getting researchers to employ sustainable practices lies more in getting the message out to researchers, rather than selling them on the benefits.

The panel rounded up with a final question about ``If you wanted to have the most disastrous effect on scientific software, what one practice would you advise people to keep doing?''
A few good responses were discussed, but the biggest laugh went to Mike Croucher who said ``Containerize everything!''

\section{Working groups} \label{sec:WGs}

After most of the lightning talks and other presentations, WSSSPE used three areas of a large room to let
attendees use sticky flip charts and sticky notes on the walls to suggest
\begin{enumerate}
\item A vision on any aspect of the work.
\item A gap or challenge
\item A project idea
\end{enumerate}
(with one area of the room for each topic.)

Once we had the walls covered in ideas, we had people organically form groups around the flip charts on the wall\footnote{Photos of the flipcharts: \url{https://github.com/WSSSPE/media/tree/master/WSSSPE4-photos}}. For those folk who worked on Vision and Gaps, after an hour, we encouraged them to finish their Vision or Gap Google Doc, and to find and join or start working on a project with a group. This led to the 12 groups that are discussed in the rest of this section, specifically:
\begin{enumerate}
\item Verifying best practices \& metrics for sustainable research software
\item Software Sustainability Alliance
\item Scientific Software Prototyping Infrastructure (S2PI)
\item Standard metadata for software (CodeMeta)
\item White paper on developing sustainable software
\item Social science for scientific software
\item Software best practices for undergraduates
\item Meaningful metrics for sustainable software
\item Coordinating access to CI for research software
\item Software engineering processes tailored for research software
\item Open research index
\item Letters of evaluation for computational scientists
\end{enumerate}

The groups worked in Google docs, and eventually created Google slides to show their work back to the larger groups. The docs and slides were stored in a common Google Drive folder\footnote{\url{https://drive.google.com/drive/u/0/folders/0BzIpChHwv-q5X3lRaWhYMXJZaUE}}. In each group's discussion, a number of Specific, Measurable, Achievable, Relevant, and Timely (SMART) steps were created, and ideally, at least the first one (to set up some communication mechanism) was completed.  In many cases, this was done through Slack (see \S\ref{sec:slack}.)


\subsection{Verifying best practices \& metrics for sustainable research software}
\label{sec:best-practices-sustainable}


Many open source projects for research software document their best practices that
contributed to the sustainability of the software.  Many open source projects also
document software metrics they use to define their research software project's success.
This group aims to aggregate several sources of these best practices for sustainability and
metrics into a consolidated list. The team then plans to create a workflow to
evaluate how open source projects stand up to these lists.

\subsubsection{Participants}

\begin{itemize}
\item Ray Idaszak <rayi@renci.org>
\item August Muench <august.muench@aas.org>
\item Jonah Miller <jmiller@perimeterinstitute.ca>
\item Lorraine Hwang <ljhwang@ucdavis.edu>
\end{itemize}

\subsubsection{Working group objective}

The overall objective of the Verifying Best Practices and Metrics for Sustainable Software working group is to take the outputs of the WSSSPE efforts to identify best practices for creation of sustainable software for science/academia, and also the outputs of WSSSPE efforts to identify metrics for sustainable software for science/academia and cross-reference these with current open source research software that successfully uses modern software engineering.  This will allow the group to identify gaps on both sides.  This approach will also allow the group to hypothesize how successful open source projects can be further improved, verify that recommended approaches for software engineering for science/academia are sufficient and valid, and that metrics for software engineering for science/academia are relevant and useful.  The group has a specific objective of getting a good cross-sampling of disparate software including community model codes, community cyberinfrastructure, community analytic tools, etc.

\subsubsection{Gap or challenge}

The gap addressed by the project is described by the following:
\begin{itemize}
\item Are the suggested software engineering for science/academia best practices verified when evaluated against open source research software projects that currently use modern software engineering?
\item Are the suggested metrics for science/academia verified to be relevant when evaluated against open source research software projects that currently use modern software engineering?
\item Do research groups have role models to follow for successful best practices in research software?
\end{itemize}

\subsubsection{Relevant people and resources}

The people spearheading this effort at the time of the writing of this workshop report are the
working group participants listed previously along with Peter Elmer <peter.elmer@cern.ch> and Hans Fangohr <fangohr@soton.ac.uk>.

Resources include research projects suggested from WSSSPE community including:
\begin{itemize}
\item FLASH (astrophysics, \url{http://flash.uchicago.edu/site/flashcode/})
\item GROMACS (chemistry, \url{http://www.gromacs.org/})
\item CIG (Earth science, \url{http://geodynamics.org/})
\item CUAHSI (Earth science, \url{https://www.cuahsi.org/})
\item CSDMS (Earth science, \url{https://csdms.colorado.edu/wiki/Main_Page})
\item OntoSoft (biology, \url{http://www.ontosoft.org/})
\item ASCL (astronomy, \url{http://ascl.net}), specifically ASCL entries sorted by citation count\footnote{\url{https://ui.adsabs.harvard.edu/\#search/q=bibstem\%3A\%22ASCL\%22&sort=citation_count\%20desc\%2C\%20bibcode\%20desc}}
\item yt-project (astronomy,  \url{http://yt-project.org})
\item Einstein Toolkit (physics, \url{https://einsteintoolkit.org/})
\item SpEC (physics, \url{https://www.black-holes.org/SpEC.html})
\item pyCLOUDY (astrophysics, \url{http://ascl.net/1304.020})
\end{itemize}

\noindent
Additional resources include these sources for best practices to be explored:

\begin{itemize}

\item Good Enough Practices in Scientific Computing, Wilson et al. 2016~\cite{DBLP:journals/corr/WilsonBCKNT16}
\begin{itemize}
\item \url{https://swcarpentry.github.io/good-enough-practices-in-scientific-computing/}
\end{itemize}

\item Best Practices for Scientific Computing, Wilson et al. 2014~\cite{bestprSC}
\begin{itemize}
\item 8 high level categories
\end{itemize}

\item Butterfly: a paradigm towards stable bio and neuro informatics tools development~\cite{10.3389/conf.fnins.2015.91.00009}

\item CIG Software Development Best Practices
\begin{itemize}
\item \url{https://geodynamics.org/cig/dev/best-practices/}
\item 6 categories; 3 Levels of detail within each: minimum, standard, target
\end{itemize}


\item Software Sustainability Institute
\begin{itemize}
\item \url{https://www.software.ac.uk}
\end{itemize}

\item WSSSPE papers from past WSSSPE workshops
\begin{itemize}
\item \url{http://wssspe.researchcomputing.org.uk/}
\item published as JORS collection (\url{http://openresearchsoftware.metajnl.com/collections/special/working-towards-sustainable-software-for-science/})
\end{itemize}

\item JOSS Peer Review and rOpenSci onboarding
\begin{itemize}
\item \url{http://joss.theoj.org/about#reviewer_guidelines}
\item Example JOSS review: \url{https://github.com/openjournals/joss-reviews/issues/43}
\item \url{https://github.com/ropensci/onboarding}
\item \url{http://ropensci.org/blog/2016/03/28/software-review}
\end{itemize}

\item RSE 2016 Talk:
\begin{itemize}
\item InterMine: Best Practices for Open Source Software (\url{http://www.rse.ac.uk/conf2016_talks#T1.1})
\end{itemize}

\item ANNIS corpus linguistic analysis tool
\begin{itemize}
\item humanities code (\url{http://corpus-tools.org/annis/})
\end{itemize}

\end{itemize}

\subsubsection{Plans}

The group's plans are for each of group members to volunteer to take on one to two projects. As an incentive, the projects that each individual takes on should be ones in which they are already interested.  This would lead to an estimated 5--10 projects to be evaluated within this project.  In terms of increasing this sample size, the group hopes that once it documents the workflow by which these evaluations are performed, others can be self-sufficient in following this workflow and contribute their own evaluations of software they wish to evaluate.

\subsubsection{SMART steps}

\begin{itemize}
\item Create mailing list for project titled:  Verify Best Practices and Metrics for Sustainable Research Software.
\item Identify where to obtain representative recommended software engineering best practices and metrics within the WSSSPE4 community including single points-of-contact.  Obtain said best practices and metrics in raw form.
\item Create GitHub site for project.
\item Identify where to obtain representative successful open source projects in some manageable number of domains that also successfully use modern software engineering.  The minimum number of projects per domain should be 1. Obtain said list of representative projects.

\begin{itemize}
\item Example sources: Astronomy source code library; high citation software papers; CIG software list; personal domain knowledge.
\end{itemize}

\item Create a case study sheet made of questions to be posed about existing research groups.
\item Perform review.
\item Aggregate results and document.
\item Document workflow of getting to these results so it can be repeated for additional domains.
\end{itemize}

\subsubsection{More information \& joining instructions}

To join this group or obtain more information about it, please send an email to the Verifying best practices \& metrics for sustainable research software group: <wssspe4-verify-best-practices@googlegroups.com>.

\subsection{Software Sustainability Alliance}
\label{sec:alliance}


The Software Sustainability Alliance working group aims to establish an alliance between organizations interested in improving the sustainability of research software. Such an alliance would ease the collaboration between member organizations to improve the sharing of expertise, resources and best practices. It is currently seeking feedback on potential member organizations, as well as the aims and scope of this alliance.

\subsubsection{Participants}

\begin{itemize}
\item Neil Chue Hong <N.ChueHong@software.ac.uk>
\item Jean Salac <jeansalac@virginia.edu>
\item Radovan Bast <radovan.bast@uit.no>
\item Lorraine Hwang <ljhwang@ucdavis.edu>
\item Karthik Ram <karthik.ram@berkeley.edu>
\item Peter Elmer <Peter.Elmer@cern.ch>
\end{itemize}

\subsubsection{Working group objective}

The overall objective of the Software Sustainability Alliance working group is to develop the steps necessary for establishing an alliance between organizations willing to engage in mutually reinforcing activities to advance the sustainability of software used in research. These organizations are funded groups or teams that aim to advance research software sustainability beyond their local university or community. More specifically, this working group aims to define the scope of this alliance and provide clear distinction with WSSSPE. The group also seeks to understand the incentives for members of this alliance and identify its key activities.

\subsubsection{Gap or challenge}

Currently, point-to-point collaboration exists between organizations, but this inadvertently results in competition or redundancy within the sustainable software community. An alliance of software sustainability organizations would ease inter-organization collaboration and the promotion of software sustainability. This alliance would also improve the pooling of competencies and the sharing of expertise. Furthermore, with the international scope of this alliance, it could support an organization who wants to hold an event in a country where another organization exists.

\subsubsection{Relevant people and resources}

The working group is looking for ideas for which organizations should be consulted or invited to join the Software Sustainability Alliance. It is also looking for feedback on the aims and scope for joining the Software Sustainability Alliance.

\subsubsection{Plans}

Throughout the scoping period, the working group will continue to refine the Software Sustainability Alliance as it collects feedback on potential members, aims and scopes. It will also update its website to reflect these changes. Next steps and further tasks will be determined based on the feedback from the scoping period.

\subsubsection{SMART steps}

The first SMART steps of the working group are outlined below:
\begin{itemize}
\item Create a draft text to explain context and incentives for joining the Software Sustainability Alliance
\item Come up with a list of potential invitees
\item Update \url{www.softwaresustainability.org} to provide information
\item Draft a letter to go out to potential alliance members
\item Create a timeline and framework for initial consultation
\end{itemize}

\subsubsection{More information \& joining instructions}

If interested, visit \url{http://softwaresustainability.org/} or email Neil Chue Hong (<N.ChueHong@software.ac.uk>) and Jean Salac (<jeansalac@virginia.edu>)

\subsection{Scientific Software Prototyping Infrastructure (S2PI)}
\label{sec:prototyping}

There is a productivity bottleneck---yet to be solved---in HPC from the human
perspective, first identified in the 1980s~\cite{barstow1982automatic}. Of all
domain-specific scientists, only a percentage of those use simulations due to
perceptions mostly related to the difficulty of using and developing those tools.
XSEDE and other resources have become a gateway for successfully increasing the
basis of HPC scientific users by facilitating their access to infrastructure and
tools~\cite{towns2014xsede}, but the development challenge remains unaddressed.
Of all scientists who use simulations, only a small percentage know how
to develop prototypes or full applications that can be later scaled by computer
scientists and engineers. Writing new parallel applications is unusual for
domain scientists and thus has created a complex dependency on specialized
scientific programmers, who are scarce~\cite{post2005computational}. However,
the capacity to quickly envision and prototype applications by domain
experts---later to be fully implemented into by scientific programmers---is critical for
acquiring maturity and proficiency in the grand challenges that Exascale
attempts to solve, and moreover for allowing scientists to develop their own
ideas quickly and productively~\cite{vinter2015prototyping}. The latter, in
addition, would increase service utilization time in HPC systems, a critical
measure of efficiency. Considering the existing code base in scientific
computing, software infrastructures that allow a clean transition from legacy
systems to new Exascale platforms while preserving flexible prototyping
towards the future are central~\cite{hwu2015transitioning}.

This group attempts to tackle the latter challenge through the development of
novel software prototyping infrastructure with an emphasis at a higher degree
of abstraction. The group deems the ability to rapidly construct software artifacts that can be
trusted in terms of the methods from their design up, easily discarded when
wrong and extended when right at low human and computational cost to be one
aspect of scientific software sustainability. In essence,
a motto for scientific software prototyping is \textit{fail hard, fail fast}.

\subsubsection{Participants}

\begin{itemize}
  \item Santiago N\'u\~nez-Corrales <nunezco2@illinois.edu>
  \item Chris Sweet <chris.sweet@nd.edu>
  \item Steven Brandt <sbrandt@cct.lsu.edu>
  \item Dominic Orchard <d.a.orchard@kent.ac.uk>
\end{itemize}

\subsubsection{Working group objective}

The group's objective is development of a prototype tool that allows domain scientists to generate
simulation prototypes through a simple, yet expressive declarative
strongly-typed programming language close to equational expressions. The result
of that specification is both an executable artifact as well as a specification
for scientific programmers to later flesh out completely and adapt to particular
infrastructures. This tool will start in the domain of computational physics,
and (probably) extend to other domains, and the code will be open source.

The underlying hypothesis of this project is that a strongly typed, declarative
problem-specification language with adequate constructs, to be implemented
through stencils and algorithmic skeletons tied to numerical methods, captures a
 subset of useful and interesting problems in physics and other domains.

\subsubsection{Gap or challenge}

The gap addressed by the project is described by
following four statements:

\begin{enumerate}
  \item Domain specific scientists need to strike a balance between being
  productive in their science areas and having some coding skills.
  \item Being a research software engineer is a career in its own right which
  has limited depth in the science basis, especially while serving large
  communities.
  \item Collaboration in science is made (mostly informally) through natural
  language and mathematics, but both present lots of opportunity for potential
  ambiguity towards software development.
  \item An unambiguous (formal) language that sets a neutral ground is missing,
  capable of lifting many of the complexities of constructing useful (and
  correct) software artifacts that may later evolve into more elaborate codes.
\end{enumerate}

The latter enumeration of elements is sustained in the fact that scientific
software packages are digital representations of partial knowledge architectures
behind research processes.

\subsubsection{Relevant people and resources}

The project is mostly self-contained and only requires development time from
participants, as well as a Git repository (active). As for skills required, the
following distribution between project participants was identified:

\begin{description}
  \item[Steven] parsing expression grammars, prototype stencil
  \item[Santiago] functional programming, scientific programming
  \item[Dominic] semantics, types, compilers, languages
  \item[Chris] numerical methods
\end{description}

In terms of practice and experience, a sample scientific community is required
in order to work around the capabilities of the tool. In addition, use cases
are required (e.g., micromagnetics~\cite{fischbacher2007systematic}).

\subsubsection{Plans}

The group's general plan is to develop a functional (meta-)prototype before WSSSPE5, and
if useful, then develop a larger software infrastructure tied to
particular cyberinfrastructure resources.

\subsubsection{SMART steps}

\begin{enumerate}
  \item Define or choose the description language depending on the potential user
  base.
  \item Develop the language interpreter/compiler for the equational description
  language.
  \item Define a set of fundamental software patterns to translate to algorithmic
  skeletons.
  \item Implement a translation mechanism from the language to the patterns.
  \item Test the system with three known solvable problems and evaluate with a
  set of functional and non-functional metrics:
  \begin{enumerate}
    \item Wave equation
    \item Newtonian evolution
    \item Boussinesq equations
  \end{enumerate}
\end{enumerate}

\subsubsection{More information \& joining instructions}

S. Brandt has shared the following background links:

\begin{itemize}
  \item Piraha parser code: \url{https://github.com/stevenrbrandt/piraha-peg/blob/master/piraha.jar}
  \item A stripped down peg file (grammar) which describes a sequence of equations:
\url{https://www.cct.lsu.edu/~sbrandt/eqns.peg}
  \item An example equation file: \url{https://www.cct.lsu.edu/~sbrandt/wave.eq}
\end{itemize}

To join this project, please visit the GitHub repository at
\url{https://github.com/nunezco2/S2PI} or join the WSSSPE Slack channel
\texttt{\#wg-sci-soft-proto}.

\subsection{CodeMeta}
\label{sec:CodeMeta}


This group, which included people who had previously been working on the CodeMeta project\footnote{\url{http://codemeta.github.io/}}, formed to address the proposed project,
 ``Define a standard for metadata to be used in current repos including authors, dependencies, \dots{}, repo URL.''

\subsubsection{Participants}

\begin{itemize}
\item Gabrielle Allen <gdallen@illinois.edu>
\item Stephan Druskat <stephan.druskat@hu-berlin.de>
\item Iain Emsley <iain.emsley@oerc.ox.ac.uk>
\item Carole Goble <carole.goble@manchester.ac.uk>
\item Chris Gwilliams <gwilliamsc@cardiff.ac.uk>
\item Rafael Jimenez <rafael.jimenez@elixir-europe.org>
\item Frank L\"{o}ffler <knarf@cct.lsu.edu>
\item Kyle Niemeyer <Kyle.Niemeyer@oregonstate.edu>
\item Thomas Robitaille <thomas.robitaille@gmail.com>
\item Rob Welch <py12rw@leeds.ac.uk>
\item Alice Allen <aallen@ascl.net>
\end{itemize}

\subsubsection{Working group objective}

The primary objective of this working group is to find ways to help the CodeMeta project come to fruition. CodeMeta seeks in part to create a ``Rosetta Stone'' for software metadata to facilitate retaining such metadata between repositories, services, registries, indexers, publishers, citation managers, and other entities that create, ingest, use, and/or store metadata about software. The project also wants to establish a JSON-LD schema as a tool for making metadata machine-readable~\cite{CodeMeta_schema}.

\subsubsection{Gap or challenge}

Research software is often not shared; that which is shared may not have much metadata associated with it, and that which does exist often does not travel further than the website on which the software resides. CodeMeta wants to incentivize software developers to release their software, encourage the development of metadata for it, enable credit assignment and citation of research software, increase its discoverability, more easily track dependencies, and enable reuse of software metadata, all goals that WSSSPE attendees have great interest in supporting.

\subsubsection{Relevant people and resources}

The CodeMeta team members (as of the WSSSPE4 meeting) are listed below; the CodeMeta project leads are shown in bold, and team members who also part of the WSSSPE working group are shown in italics :

\begin{itemize}
\item {\bf Carl Boettiger}, UC Berkeley
\item {\bf Matt Jones}, NCEAS
\item {\bf Arfon Smith}, Space Telescope Science Institute
\item {\bf Abby Mayes}, Mozilla Science Lab
\item Yolanda Gil, USC ISI
\item Peter Slaughter, NCEAS
\item Patricia Cruse, DataCite
\item Neil Chue Hong SSI
\item Merc\`{e} Crosas, Harvard IQSS
\item Martin Fenner, DataCite
\item Mark Hahnel, figshare
\item Luke Coy, RIT \& MSL
\item {\em Kyle Niemeyer}, Oregon State
\item Krzysztof Nowak, Zenodo
\item Daniel S. Katz, NCSA
\item {\em Carole Goble}, University of Manchester
\item Ashley Sands, UCLA
\item {\em Alice Allen}, ASCL

\end {itemize}

Resources already established by the CodeMeta team can be found online:

\begin{itemize}
\item Main website and meeting information (\url{http://codemeta.github.io/})
\item Github repository (\url{https://github.com/codemeta/codemeta})
\item List of milestones (\url{https://github.com/codemeta/codemeta/milestones})
\item Gitter discussion site (\url{https://gitter.im/codemeta/codemeta})
\end {itemize}

Other resources to draw on are the software metadata vocabularies of Schema.org\footnote{\url{https://schema.org/SoftwareSourceCode}} and DOAP\footnote{\url{https://github.com/ewilderj/doap}}, and the FORCE11 Software Citation Principles~\cite{Smith:2016sc}. Relevant documents on research software, including the Guidelines for persistently identifying software using DataCite~\cite{Gent2015} are available on the UK's Research Data Network's Research Software web page\footnote{\url{https://research-data-network.readme.io/docs/research-software}}.

\subsubsection{Plans}

The working group plans to engage with those already working on CodeMeta, to examine the existing CodeMeta crosswalk table to see what improvements and additions might be made, and to determine how to engage the community and provide ongoing social engagement and structure. Further, the group wants to assist in the implementation of the specifications and, by providing an outsider's view, contribute suggestions for more understandable project documentation. Finally, the WG seeks a better way or ways to present the crosswalk table to make it more easily understood and consumable by research software communities.

\subsubsection{SMART steps}

With a comparatively large working group, the use of the established CodeMeta Github repository's issue tracking, and the quick responsiveness of CodeMeta lead Matt Jones to a barrage of questions, comments, and logged issues, several of the group's SMART steps were completed at WSSSPE; these are identified below in italics.

\begin{itemize}
\item {\em Write to the managers of CodeMeta project to start dialog with the two groups}
\item {\em Post questions generated by our discussion as issues on the CodeMeta repository}
\item Look at the crosswalk table and Schema.org data elements for common elements
\item Identify text in the project documentation that is unclear and suggest changes
\item Define a list of services to be added to the crosswalk table and match terms
\item {\em Identify two roles and information on CodeMeta that would be useful to these people for engaging with the project}
\item Determine a better way or ways to present the crosswalk table
\item Create a mailing list for WSSSPE CodeMeta participants
\end{itemize}

\subsubsection{More information \& joining instructions}

Notes for the working group were taken in real time in a Google document\footnote{\url{https://docs.google.com/document/d/1UxlHIoBRgVWB8NAXYf4Q0yS7PAqa-EYwFfsNuTKSPbs/edit}} and include email replies in response to WG questions from one of the CodeMeta project leads.

Because of the work done at WSSSPE4, the CodeMeta project README file\footnote{\url{https://github.com/codemeta/codemeta/blob/master/README.md}} was greatly expanded to include a description of the project that is geared to those with little or no prior knowledge of the project, a list of contributors, information on how one can get involved, a brief project history and who is managing the project, and links to additional information. Though a Google group mailing list has been established for the working group, the easiest way to engage with the CodeMeta project is through its Github repository\footnote{\url{https://github.com/codemeta/codemeta}}.

\subsection{White paper on developing sustainable software}
\label{sec:best-practices-developing}



Many diverse aspects and dimensions of developing sustainable software can be investigated,
such as economic, technical, environmental, and social. This group aims to write white papers
that will focus on scientific environments and their implications, targeted at developers
and project managers of scientific software. Given the complexity of this field, it is important to select
a subset of sustainability aspects for the white papers. The idea is to create a series of papers instead of trying
to tackle all important topics in one paper. For the first white paper, the group aims to set the
stage with successful use cases and an analysis of why they have been successful.
Further topics will include community-related practices, government and management, funding,
metrics, tools, and usability. The group will collect feedback from the WSSSPE
community on this first white paper and extend it to a journal paper.

\subsubsection{Participants}

\begin{itemize}
\item Jeffrey C. Carver <carver@cs.ua.edu>
\item Neil Chue Hong <N.ChueHong@software.ac.uk>
\item Tom Crick <tcrick@cardiffmet.ac.uk>
\item Miguel de~Val-Borro <valborro@princeton.edu>
\item Hans Fangohr <fangohr@soton.ac.uk>
\item Sandra Gesing <sandra.gesing@nd.edu>
\item Derek Groen <Derek.Groen@brunel.ac.uk>
\item Dan Gunter <DKGunter@lbl.gov>
\item Daniel S. Katz <d.katz@ieee.org>
\item Alexander Konovalov <alexander.konovalov@st-andrews.ac.uk>
\item Frank L\"offler <knarf@cct.lsu.edu>
\item Suresh Marru <smarru@iu.edu>
\item Kyle E.~Niemeyer <kyle.niemeyer@oregonstate.edu>
\item Abani Patra <abani@buffalo.edu>
\item Francisco Queiroz <chico@puc-rio.br>
\end{itemize}

\subsubsection{Working group objective}


The working group aims to start a series of papers and supporting developers and project managers of scientific software. While there are already a few papers available on best practices and sustainability of scientific software in general, the group's goal is to create a series of papers that lead to consensus in the community, tackle many diverse aspects, and stay up-to-date with new trends. This goal is quite ambitious, and the group hopes to attract more authors over time who would like to contribute to specific aspects and take the lead on them.

\subsubsection{Gap or challenge}

The challenge for the group is to accomplish a first version of the white paper to kick off the series.
The complexity of the topic has led to a first attempt in 2016 that may have been a bit too ambitious,
by trying to cover a wide range of topics. Addressing different topics over time seems to
be a better strategy to finalize a paper. The series would fill also a gap on best practices to which the community can contribute.

\subsubsection{Relevant people and resources}

The list of contributors, as of 10 November 2016, is the same as the list of participants above.

\subsubsection{Plans}

The resulting draft white paper will be distributed via the WSSSPE email list. After collecting feedback on it, the group's plans are to attract a wider community to contribute to an extended journal paper version and to investigate whether more people would like to be involved in the white paper series.

\subsubsection{SMART steps}

First eight SMART steps proposed: \\
\begin{itemize}
\item List existing contributors and open GitHub for new ones -- Sandra Gesing (done)
\item Establish preliminary timeline  -- Sandra Gesing (done)
\item Organize related work -- Francisco Queiroz (done)
\item Define scope of the paper -- Sandra Gesing (done)
\item Suggest new sections -- Francisco Queiroz (done)
\item Define leading authors and contributors for sections -- Francisco Queiroz (done)
\item Finalize first version -- Sandra Gesing (in progress)
\item Distribute to WSSSPE community and collect feedback -- Sandra Gesing (in progress)
\end{itemize}

\subsubsection{More information \& joining instructions}

The GitHub repository for the white paper can be found at \url{https://github.com/WSSSPE/WG-Best-Practices}. For more information and requests, join the \texttt{\#wg-best-practices} channel at WSSSPE's Slack team.

\subsection{Social science for scientific software}
\label{sec:social}



This working group is motivated by the goal of building better connections between academic researchers who are studying topics in or relating to software sustainability with practitioners, managers, and administrators who are working in the area of software sustainability. In any domain, bringing research and practice closer together a mutually beneficial goal, but also has its challenges. This group met to discuss existing research projects and findings that might be relevant for RSEs and others in the software sustainability domain, then identified several gaps and challenges to tackle in bringing research and practice together.

\subsubsection{Participants}

\begin{itemize}
  \item Stuart Geiger <stuart@stuartgeiger.com>
  \item Lorraine Hwang <ljhwang@ucdavis.edu>
  \item Robert McDonald <rhmcdona@indiana.edu>
\end{itemize}

\subsubsection{Working group objective}

To give social scientists and practitioners working on scientific software, software sustainability, and open source and a better understanding of each other's work, as well as help them connect and coordinate on specific projects of mutual interest.

\subsubsection{Gap or challenge}

There is more and more academic research being done on topics related to software sustainability, including work on software engineering practices and management of open source projects. However, academic research in general is often siloed for many reasons, and work on topics relevant to software sustainability is no exception. There are many academic studies and projects which may be relevant for practitioners working in this area, and many projects and initiatives by practitioners that may be relevant for social scientists. However, there is a gap between these two research and practice. Furthermore, the group also recognize that academic social science and research software engineering are not monolithic, and there is a need to connect people who care about research-driven best practices inside of these two domains with each other as well.

\subsubsection{Relevant people and resources}

People: a network of invested social scientists and research software engineers who care about research on software engineering practices
Resources: a mailing list and a central repository for collecting literature and documenting lessons learned

\subsubsection{Plans}

The working group plans to briefly survey existing literature to get a better sense of the academic research landscape, facilitate some initial dialog between academic researchers and practitioners at WSSSPE4, then identify needed actions that would be mutually beneficial to researchers and practitioners. This plan resulted in the creation of the following SMART steps listed in the next section.

\subsubsection{SMART steps}

\begin{itemize}
\item Collect research questions RSEs have for social scientists (done)
\item Create a mailing list for people interested in this topic (done) \url{https://groups.google.com/forum/#!forum/researchsoftwarestudies}
\item Create a central repository for collecting literature and documenting lessons learned
\item Publicize mailing list and collect (more) research questions/topics/literature/researchers
\item Write up synthesis document 
\end{itemize}

\subsubsection{More information \& joining instructions}

Join the mailing list at \url{https://groups.google.com/forum/#!forum/researchsoftwarestudies}

\subsection{Software best practices for undergraduates}
\label{sec:best-practices-undergrads}



This working group was motivated by the perceived prevalence of
so-called ``hidden code'' in scientific communities: code written by
individual researchers in an unsustainable way that is never shared
with the larger community. Participants recalled their own experience
working with colleagues who write hidden code or inheriting
unsustainable hidden code from collaborators.

The question is, then, how to prevent researchers from writing hidden
code? The participants hypothesized that the best strategy is to catch
researchers while they are still in training and teach them software
best-practices. Therefore, this working group was formed with the goal
of developing courses on software best practices aimed at
undergraduate students studying domain science. The program might be
similar to a Software Carpentry or Data Carpentry workshop but with a
focus for domain scientists.

\subsubsection{Participants}

\begin{itemize}
  \item Jonah Miller <jmiller@perimeterinstitute.ca>
  \item Aleksandra Nenadic <a.nenadic@manchester.ac.uk>
  \item Raniere Silva <raniere.silva@software.ac.uk>
  \item Francisco Queiroz <chico@puc-rio.br>
  \item Hans Fangohr <fangohr@soton.ac.uk>
  \item Prabhjyot Sing <prabhjyot10@gmail.com>
\end{itemize}

\noindent Aleksandra Nenadic and Raniere Silva are involved
in Software and Data Carpentry.

\subsubsection{Working group objective}

Implementation of a course aimed at each domain science is a long term
goal. However, a short-term, achievable goal, is to develop a
curriculum for a course aimed at a single domain science. Since the
participants in the working group have expertise in physics, this is a
natural target.

\subsubsection{Gap or challenge}

The challenge is that people are writing unsustainable scientific
software without ever learning that there is a better way. Here we
focus on training in good software design and engineering in this
context.

\subsubsection{Relevant people and resources}

The development of a successful curriculum relies on the expertise of
software engineers to describe the best practices, domain experts to
describe model problems and work-flow, and instructors to formulate
the pedagogy. Ideally these people will be brought together for a
short workshop or hackathon with the goal of drafting the
curriculum. Later, organizational partners will be required to
actually implement the program.

\subsubsection{Plans}

The working group will seek funding to organize a hackathon to develop
the curriculum and seek the required expertise to invite. The
hackathon will be implemented and a draft of the curriculum will be
written within the next year.

\subsubsection{SMART steps}

The working group will accomplish or has accomplished the following
smart steps:

\begin{enumerate}
\item Create channels of communication including a mailing list, a
  slack channel, and a git repository (done)
\item Review literature on best and good-enough practices for
  scientific computing as well as literature on pedagogy for
  scientific and research computing (by February 1, 2017)
\item Organize a telecon to organize a workshop to write the
  curriculum and decide how to bring in outside experts (by February 1,
  2017)
\item Write a draft of the curriculum (by April 1, 2017)
\item Invite the broader community to contribute to a more complete
  curriculum (by April 30, 2017)
\end{enumerate}

\subsubsection{More information \& joining instructions}

Anyone interested in contributing should get in touch via one of the
following methods:

\begin{itemize}
\item Join our Google Group/mailing list
  \cite{WSSSPEUndergradGoogleGroup}.
\item Join the WSSSPE Slack (\S\ref{sec:slack}) and the
  \texttt{\#wg-undergraduatecourse} channel.
\item Ask to join the WSSSPE4-undergraduate-course organization
  \cite{WSSSPEUndergradGithub}.
\end{itemize}

\subsection{Meaningful metrics for sustainable software}
\label{sec:metrics}



Meaningful Metrics for Sustainable Scientific Software aims to increase the visibility of the quality of scientific software, facilitate the reusability of scientific software, and promote the best software practices by standardizing metrics via interviews with scientific software developers. This working group believes improving the current software metrics system will increase software sustainability. Currently, there are inefficiencies regarding software duplication, sustainability, and selection, as well as others, within the scientific software community. In order to address these inefficiencies, Meaningful Metrics for Sustainable Scientific Software aims to create a goal-oriented method to collecting productive metrics by focusing on the developer side of software.

\subsubsection{Participants}


\begin{itemize}
\item Emily Chen <echen35@illinois.edu>
\item Patricia Lago <pdotlago@gmail.com>
\item Udit Nangia <unangi2@illinois.edu>
\item Tengyu Ma <tengyuma10717@gmail.com>
\item Aseel Aldabjan <a.dabjan@hotmail.com>
\end{itemize}

\subsubsection{Working group objective}


In the context of the larger objective, Meaningful Metrics for Sustainable Scientific Software hopes to find an efficient solution for streamlining the process of collecting and utilizing metrics to benefit software sustainability, as well as minimize the current inefficiencies within the scientific software community. Finding meaningful metrics will improve software evaluation and comparison, thus reducing the effort spent on seeking scientific software.

\subsubsection{Gap or challenge}


The gap being addressed is the lack of a standard for collecting and presenting metrics. This gap delays workflow and creates a multitude of tedious tasks, including searching for the best-fit software, unknowingly duplicating software, and other related busy-work. The need for metric standardization stems from the abundance of scientific, both ``dark'' and open source, software and the difficulties of ensuring the software is sustainable. 

\subsubsection{Relevant people and resources}


Relevant people and resources include scientific software developers, researchers that use scientific software, scientific software funding institutions.

\subsubsection{Plans}


This working group plans on interviewing scientific software developers to form metrics from the goals they have for their software.

\subsubsection{SMART steps}


The SMART steps Meaningful Metrics for Scientific Software proposed are:
\begin{enumerate}
\item Identify the population of scientific software developers who are willing to be interviewed
\item Define the interview questions and organize them into categories
\item Interview the participants and map the survey results to goals
\item Convert the goals to feasible, meaningful metrics
\item Analyze the collected metrics
\end{enumerate}

\subsubsection{More information \& joining instructions}


For more information, please contact Emily Chen at <echen35@illinois.edu>.

\subsection{Coordinating access to continuous integration (CI) for research software}
\label{sec:access}


Each developer of software with uncommon needs (hardware, software, libraries, data sets), non-public code, or tests that exceed time limits for free plans must acquire, setup, and maintain their own continuous integration systems because their needs make them ineligible for popular free services such as Travis CI.  For example, software groups that develop BIOUNO, CI4SI, or GROMACS have done this.

Note that Debian provides a testing infrastructure (\url{https://ci.debian.net/doc/}) that is mature and supports hardware and other requirement specifications.

\subsubsection{Participants}

\begin{itemize}
  \item Xinlian Liu <liu@hood.edu>
  \item Mark Abraham <mjab@kth.se>
  \item Sameer Shende <sameer@cs.uoregon.edu>
  \item James Hetherington <j.hetherington@ucl.ac.uk>
  \item Dominik Kempf <dominic.kempf@iwr.uni-heidelberg.de>
  \item Michael R. Crusoe <michael.crusoe@gmail.com>
  \item Radovan Bast <radovan.bast@uit.no>
\end{itemize}

\subsubsection{Working group objective}

This group is interested in reducing the burden of different projects having to build and maintain their own continuous integration systems (when publicly available CI are not a fit), by coordinating and sharing this burden across multiple projects.

\subsubsection{Gap or challenge}

The scope of the group's interest is any type of testing, though interactive access for troubleshooting would be of particular interest beyond just automated testing.

However, there are a lot of open issues:

\begin{itemize}
\item Some similar work has already been done.  How can this group apply and/or learn from that work, rather than reinventing the wheel?
\item How to ensure this will work across disciplines?
\item Meaningful CI for large projects may need hundreds of CPU hours per day
\end{itemize}

\subsubsection{Relevant people and resources}

This would need to be picked up by a set of projects, potentially both small and large.

Possibly working with the RSE community would be useful.

\subsubsection{Plans}

Some possible goals include:
\begin{itemize}
\item Acquire additional hardware such as GPUs, Xeon PHI, FPGAs and add/share them to e.g. Debian's testing infrastructure or to a shared Jenkins-based infrastructure
\item Extending Debian's scope to include published but not mature software
\item Implementing the same interface but with specialized hardware or available to non-public codes.
\item \url{https://reproducible-builds.org} but for CI
\end{itemize}

\subsubsection{SMART steps}

\begin{enumerate}
\item Learn how to donate machines to Debian
\item Find funding -- NeIC (\url{https://neic.nordforsk.org}) through the CodeRefinery project (\url{http://coderefinery.org}) is building a CI infrastructure (in progress) that will be provided to selected projects (through an open call)
\item Build community
\end{enumerate}

\subsubsection{More information \& joining instructions}

For more information, see \url{https://groups.google.com/forum/#!forum/continuous-integration-for-research-software}


\subsection{Software engineering processes tailored for research software}
\label{sec:soft-eng}


This working group is concerned with identifying processes that are not adequately
covered by general software engineering. Verification and testing is
the first subtopic that it addresses.
Computational science and engineering applications have many moving
parts that need to interoperate with one another. The accuracy and
reliability of results produced by scientific software depends not
only on the individual components behaving correctly, but also on the
validity of their interactions.
As scientific understanding grows,
the corresponding computational software models are refined, leading
to more complex codes. Increasing complexity makes them more prone to
defects, not only in individual code units, but also in interaction
among units. Therefore, a strong verification process combined with a
rigorous testing regime plays a critical role in the prevention of
generating incorrect scientific results. However, most science teams
struggle to find a good solution for themselves. Causes range from
lack of exposure to the practices, to distrust of adopting practices
because they do not meet the needs of the teams developing such
software.  The current focus of our effort is on testing because it is the first step
towards building a software processes that can lead to provenance and
reproducibility, the hallmarks of quality science.

\subsubsection{Participants}
\begin{itemize}
\item Mark Abraham <mjab@kth.se>
\item Anshu Dubey <adubey@anl.gov>
\item Hans Fangohr <fangohr@soton.ac.uk>
\item Dominic Kempf <dominic.kempf@iwr.uni-heidelberg.de>
\item Eric Seidel <eseidel@cs.ucsd.edu>
\end{itemize}

\subsubsection{Working group objective}
Scientific computing software lags behind commercial software in
adoption of software engineering practices. This
gap is particularly acute in the area of software testing,
verification, and validation, where the standard practices are
simultaneously inadequate and overly onerous.
This group aims to close the gap.

\subsubsection{Gap or challenge}
Computational science code developers often lack of exposure to
regular testing and its benefits. Good developers will test their code to
verify that it operates as expected; however, they may not appreciate
that without regular testing, defects can be introduced
inadvertently. An even bigger challenge is that those who understand
the importance of regular testing do not often find much help from
software engineering literature. There is a significant gap between
the testing gospel and its applicability to computational science. This
gap leads to frustration and abandonment of the good with the bad. Some
relevant literature exists, in particular experiences from
practitioners in computational science who developed their own
solutions. However, this literature is scattered among many different
forums, and can be challenging to find. This working group aims to
address this gap by curating the existing content and contributing
content where none exists.

\subsubsection{Relevant people and resources}

The working group will benefit from wide participation by developers
of large computational science codes. The reason is that the
management of such codes becomes intractable without adopting some
software process and a testing regime. The experiences and
customizations vary, and the community will benefit from hearing about
as many as possible. The seed resources required are fairly
minimal. The group has started a git repository for collecting the
existing references. That, and a few volunteers reading through the
references is all the group needs in the beginning. As it gathers more
knowledge and pinpoint gaps, it may need more resources to reach out
to a wider group of developers to collect more information.

\subsubsection{Plans}

This working group will (1) conduct a literature survey to gauge the extent
of awareness of the issue in general, (2) generate content useful for
the community where needed, and (3) curate the collected and added
content for the use of the community.

\subsubsection{SMART steps}

Our first few SMART steps are :
\begin{itemize}
\item Create a channel in wssspe.slack.com -- {\em done}
\item Create github repository
  {\url{http://github.com/wssspe/testing-in-science}} -- {\em done}
\item Find and gather existing publications in repository -- {\em ongoing}
\item Review and summarize material. Decide whether we consider this
  sufficient. If yes, then the group will put a brief report together and
  conclude the working group.
\item If the group does not consider the material adequate, it will start research
  and gather methodologies for testing in science
\item Write a document accessible to computational science and
  software engineering community, and publish document at a citable forum.
\end{itemize}

\subsubsection{More information \& joining instructions}

Readers interested in getting more information should get in touch
with a member of the working group. The working group
has a channel in the WSSSPE Slack. The channel is called
\texttt{\#wg-testing-in-science}.
(See \S\ref{sec:slack} for instructions on how to join the Slack WSSSPE team.)
Additionally, a git
repository (\url{https://github.com/WSSSPE/WG-Best-Practices.git}) exists for contributing content and reference to, and
curation of the existing literature on this topic.

\subsection{Open research index}
\label{sec:open-research-index}


The aim of this group is to investigate the building of an index of research products in an open sustainable manner.  Its goal is not to eliminate commercial products, but to build on what is there and provide data and services that are missing.

The Open Research Index should take in all research products (papers, software, datasets, workflows, etc.) from their publishers and recorders (journals, societies, domain repositories, government [open access] repositories, preprint servers, general repositories [e.g., figshare, zenodo]) and other services (CrossRef, ORCID).
Each product should list authors and citations and allow people to search the resulting network.
Users should also be able to interact with their own record and edit it, like Google Scholar allows.

\subsubsection{Participants}

\begin{itemize}
\item Gabrielle Allen <gdallen@illinois.edu>
\item Bruce Childers <childers@pitt.edu>
\item Robert Haines <robert.haines@manchester.ac.uk>
\item Caroline Jay <caroline.jay@manchester.ac.uk>
\item Daniel S. Katz <d.katz@ieee.org>
\item Robert McDonald <rhmcdona@indiana.edu>
\item Daniel Mosse <mosse@cs.pitt.edu>
\item Kyle Niemeyer <Kyle.Niemeyer@oregonstate.edu>
\end{itemize}

\subsubsection{Working group objective}

The working group's plans are relatively simple to express, though quite complex to undertake:

\begin{enumerate}
\item Determine a plan to build an open research index that allows various stakeholders to satisfy their needs
\item Then determine if the plan is feasible
\item If so, then obtain resources
\item If successful, then build the index
\end{enumerate}

\subsubsection{Gap or challenge}

Google and others provide some services now.
But these services (and the underlying data) could be removed at any time.
And the community cannot build new services.

\subsubsection{Relevant people and resources}

People/organizations who might be willing to contribute (or whose expertise is needed), and how they will contribute:

\begin{itemize}
\item ORCID, CrossRef, DataCite, ImpactStory/Depsy, \url{altmetrics.org}, Plum Analytics, GitHub, Open AIRE, CHORUS, FORCE11, COS, SHARE, CASRAI, Portico, \url{softwareheritage.org}, DBLP, eSTEP (CWO)
\item We could work with some initial publishers who don't have competing services (domain societies), PubMed
\item We need to discuss this idea with potential funders, and ask what data/services would they like to have?  And are they willing to invest in this?
\end{itemize}

The external resources that are needed are money and time; it is unclear how they can be brought in.

\subsubsection{Plans}

At the time of the meeting, and today as well, it is unclear who has the time and energy to pursue this idea.

If we identify a leader (PI), a plan could be:
\begin{enumerate}
\item Obtain funding for a set of initial discussions (or find a group who feel that this is important enough that they will take the time to do it anyhow)
\item Talk with potential partners (publishers, orgs, funders) (could be combined with meeting below)
\item Talk with the Google Scholar developer for better understanding of what they did and are planning in the future
\item Obtain funding for a meeting (perhaps from NSF, perhaps a Dagstuhl meeting)
\item Hold a community meeting to define a plan (which should be large enough to include representative of all stakeholder groups, and small enough that we still have a good open discussion)
\item Obtain funding to implement the plan, including contributions and commitments from the stakeholders
\end{enumerate}

Alternatively or additionally, the group could build a mailing list for us and discuss further, depending on how receptive others are to this idea.  At WSSSPE4, there seemed to be enough interest to do this, so the group set up a Slack channel within the WSSSPE team, called \texttt{\#wg-open-research-idx}. (See \S\ref{sec:slack} for instructions on how to join the Slack WSSSPE team.)

We also discussed some possible more detailed plans:

\begin{enumerate}
\item Look at curriculum lattes (Brazil) and/or Researchfish (UK).  They have some problems that people do not like, the group could check what features they have and what the issues are.

\item Look at what CrossRef data is

\item What research products should the index track?

\item What model should be used?
\begin{itemize}
    \item Crawling and obtaining data
    \item Consortium model
    \item Need analysis and decision
\end{itemize}

\item Need agreements with publishers, repositories, services
\begin{itemize}
    \item Some of these will have costs, though some may be free
    \item Need to determine costs
    \item Perhaps publishers would pay us to be listed eventually, as we gain power
    \item To some extent, the group would be competing with some of the publishers' products
\end{itemize}

\item If crawling model:
\begin{itemize}
    \item How to crawl all of these products?
    \item What data (metadata) should be the output of the crawls? What happens when the products being crawled and/or their metadata change?
    \item How to store all of this data (metadata)?
\end{itemize}

\item What services to provide on this data (metadata)?

\item Who will create these services?

\item What services will the publishers allow to be provided?

\item Need to develop a plan that shows incremental progress and successes

\item Consider applying to Google project call to get Google interest and support?
\end{enumerate}

\subsubsection{SMART steps}

The working group discussed a number of SMART-like steps, though most of them did not fulfill all the SMART criteria, such as identifying who would do them and when:

\begin{enumerate}
\item Develop a community mailing list/forum (Kyle, during the workshop).  This was completed; see \S\ref{sec:wg-open-research-index-list}.

\item Find a PI (we suggest Neil Chue Hong)

\item Apply for initial funding. For the US, the group discussed the Arnold Foundation Open Science award as a planning activity (who, due Dec 15), though without a PI, this seems unlikely to be done.
For the UK, there wasn't a specific suggestion.

\item Determine project name, initial vision, initial mission, logo (who, when)

\item Identify initial use cases (who, when)

\item Identify potential stakeholders and partners (who, when)

\item Discuss this idea with potential partners (who, when).
Then get their feedback \& buy-in.
While doing so, seek suggestions for the right project manager to coordinate this

\item Build up an advisory board (who, when)

\item Build up a leadership committee and determine a leader (who, when)

\item Hire a project manager (who, when)

\item Iterate use case with stakeholders (who, when),
e.g., funders, Researchfish, researchers ?

\item Iterate vision \& mission with leadership committee and advisory board (who, when)

\item Develop initial development plan (who, when)

\end{enumerate}

\subsubsection{More information \& joining instructions}\label{sec:wg-open-research-index-list}

If you are interested in more information, please join the WSSSPE team's Slack channel for this working group: \texttt{\#wg-open-research-idx}. (See \S\ref{sec:slack} for instructions on how to join the Slack WSSSPE team.)

\subsection{Letters of evaluation for computational scientists}
\label{sec:letters}


Letters of reference used for hiring, tenure, and promotion purposes
are typically first read by departmental committees that have
expertise in the core areas represented by the department. On the
other hand, computational scientists and in particular researchers
working on scientific software -- more or less by definition --
are typically interdisciplinary, and the achievements that are
assessed in such letters are consequently often meaningless or at
least hard to evaluate for disciplinary committees.

Both committees and letter writers need to be aware of this. Providing
best practice examples, and training both writers and recipients of
letters may be necessary to level the playing field for computational
scientists.

\subsubsection{Participants}

\begin{itemize}
\item Alice Allen <aallen@ascl.net>
\item Gabrielle Allen <gdallen@illinois.edu>
\item Wolfgang Bangerth <bangerth@colostate.edu>
\item Bruce Childers <childers@cs.pitt.edu>
\item Lorraine Hwang <ljhwang@ucdavis.edu>
\item Daniel S. Katz <d.katz@ieee.org>
\item Frank L{\"o}ffler <knarf@cct.lsu.edu>
\item Kyle Niemeyer <Kyle.Niemeyer@oregonstate.edu>
\item Janos Sallai <janos.sallai@vanderbilt.edu>
\end{itemize}

\subsubsection{Working group objective}

Scientists working on scientific software are often located in
disciplinary departments, depending on whether their software
originates from the mathematical, physical, chemical, or other
disciplines. As a consequence, they are frequently outside the core
areas of their science, and their contributions are typically to both
the research activity their software enables, as well as on
algorithm and implementation aspects. This presents issues when
letters of evaluation for hiring, tenure, and promotion do not specifically
cover how this is relevant to the discipline.%
\footnote{The extensive use of such letters, and the problems that are
  associated with it, may be an issue specific to the United States.}

This group believes that the authors of scientific software provide important
services to departments that are no less than strictly disciplinary
research. Consequently, leveling the playing field with more
disciplinary candidates for hires, tenure, and promotion requires that
letter writers be aware of how their letters will be read by
committees. It also requires that committees be aware that such
letters often look different and may provide a different perspective
of how a candidate's achievements should be assessed. For example, in
mathematics a typical candidate would be evaluated on the difficulty
and depth of the statements she may have proven in their papers. On
the other hand, an author of mathematical software would likely be
evaluated based on the impact of her software, or the number of citations of
the publications that describe it. She may also be evaluated by how
widely the software is used \textit{outside} mathematics, a criterion
that is rarely used for more disciplinary mathematicians.

\subsubsection{Gap or challenge}

Addressing this problem likely requires that letter writers pay
particular attention to who exactly the audience of such letters is,
and tailor the message by specifically highlighting how the work of a
candidate benefits the department and discipline that is considering a
candidate. On the other hand, committee members also need to pay
attention to the fact that there are areas that are important to the mission of
their department and professional community in which different
standards for evaluation hold.

\subsubsection{Relevant people and resources}

Affecting letter writers and readers essentially requires raising
awareness of the issue. Various organizations have attempted to do so
through letters to the community, adjusting definitions of what counts
in science, etc. Specifically, the group is aware of the following
resources:

\begin{itemize}
\item The Computing Research Association has produced a ``Best
  practices memo'' on the issue \cite{PSU99}. Among other points,
  the memo also clarified that conference proceedings are
  \textit{the} venue in Computer Science, and this has been used to
  convince deans and provosts of the value of these proceedings
  publications. The article may serve as a template for computational
  science.
\item The National Science Foundation has produced a number of ``Dear
  Colleague'' letter that consider the issue of software and, taken
  together, define a ``product'' for the purposes of
  2-page biosketches in a way that does not include only publications, but also
  software.%
  \footnote{The most pertinent such letter is NSF
      14-059, ``Dear Colleague Letter - Supporting Scientific
      Discovery through Norms and Practices for Software and Data
      Citation and Attribution''~\cite{nsf-dcl-citation}.}
  This definition is ultimately codified in the National Science
  Foundation's ``Proposal and Award Policies and Procedures Guide''
  (often abbreviated as GPG) which in 2013 renamed the
  ``Publications'' section of 2-page biosketches to ``Products'' (see
  Chapter II.C.2.f(i)(c)) and now explicitly allows software to be
  referenced in this section.
\item There is also a 1995 report by the National Academies about
  Computer Science that addresses some of these issues, specifically
  on how writing software should be considered in comparison to more
  theoretical research \cite{NRC-careers-94}.
\end{itemize}

\subsubsection{Plans}

There are essentially two important strategies that build on each
other: (i) raising awareness of the problem beyond just those who are
affected by it (namely, computational scientists working on scientific
software), and (ii) providing letter writers, letter readers, and
evaluating committees with guidance on what criteria are relevant in
assessing scientific software authors.

Concrete guidance---for example in the form of ``Dear Colleague''
letters as mentioned in the previous section---is most valuable if it
comes from respected bodies such as professional societies or
established and respected organizations. Getting these to act
will only be possible if the problem is widely acknowledged, and so
the first of the goals above should be the current focus.

The group will attempt to address it by organizing sessions at conferences
that address the career problem, as well as writing editorials that
can be published in the magazines of professional societies. These
editorials ought to outline best practices for letter writers that
specifically (i) make clear the contribution of a candidate to their
interdisciplinary area, (ii) the relevance to their home discipline, and
(iii) why writing software is good for the discipline itself.

\subsubsection{SMART steps}

The group has identified a few next steps, even if they may not satisfy the
exact criteria of the SMART system. Specifically:
\begin{itemize}
  \item Find and review the documents listed above, send out summary
    (Wolfgang Bangerth)
  \item Gather feedback from the other members of the group
  \item Write editorial for SIAM (Wolfgang)
  \item Write two emails to each other group members listed above about
    also writing editorials in their communities.
\end{itemize}
In the longer term, the group hopes to gather others in the home disciplines of its members to
let professional organizations weigh in.

\subsubsection{More information \& joining instructions}

The section on resources above lists a number of letters and reports
that are worth reading. Ultimately, building a community large enough
to affect change is important; contact Wolfgang Bangerth at
<bangerth@colostate.edu> if you are interested in helping.

\section{Slack team and channels}\label{sec:slack}

As part of the discussion at WSSSPE4, we created a Slack team for WSSSPE~\cite{WSSSPESlack}.  To join it, go to \url{https://wssspe.signup.team/}.  A number of the working groups also created channels within the team; see the subsections in \S\ref{sec:WGs} for the names of these channels.

\section{Attendee survey \label{sec:survey}}


All participants present on the last day of the meeting, 14 September 2016, were asked to complete the online survey administered through Qualtrics\footnote{\url{https://www.qualtrics.com}.}.
44 responses were recorded, with one respondent indicating that they took the survey twice.
The survey URL was also later distributed to attendees by email and advertised on twitter for attendees who were not there on the last day to have a chance to respond, and one additional response was recorded after the meeting.
Hence, the total number of unique individuals responding to the survey is likely 44.
The survey contained nine multiple choice and open-text answer questions.
Text responses have been alphabetize to preserve anonymity.
The survey had a 100\% completion rate.
Appendix~\ref{sec:survey_details} contains the detailed survey results.

In general, the respondents were highly satisfied with the meeting and interested in continuing and participating in WSSSPE activities.
The mix of topics and types of interactions---talks, panels and discussion---was well balanced with several indicating a desire for time for questions and answers following talks as well for discussions in general.
Many respondents indicated that they will remain engaged in WSSSPE working group initiatives.
However, demand for a professional organization encompassing WSSSPE interests was weaker with over half interested in joining and a large percentage willing to consider the idea.
Respondents were grateful for the opportunity to network, explore new collaborations and for travel support to the conference.

The survey results indicate that future WSSSPE conferences should consider the balance of the attendees.
This year's meeting had many first time and early career participants who would have benefited from a review of basic concepts and terminology including what it means for software to be sustainable and the roles of research software engineers.
Additional emphasis and topics to explore would be the inclusion of more case studies, focus on the decision making process in developing and using software, deeper dives into selected topics, tutorials, software training within and outside of STEM, lowering the barriers to implementing best practices in software development, and progress towards executing WSSSPE's vision.

\section{Conclusions} \label{sec:conclusions}

In WSSSPE4, we heard about a number of interesting projects and ideas, and used those
ideas to create and form working groups, intended to start at the workshop and then
continue afterwards, to address challenges that arose from the workshop ideas.  This workshop
has reinforced the lesson from WSSSPE3 that it is relatively easy to get motivated people
to attend a meeting and productively spend their time there both doing work and planning
more work, but it is very hard to get that additional work after the meeting to take place.
The main problem seems to be one of time.  Once the attendees have agree to spend
their time at the workshop, they put their energy into doing so productively, but they have not
really committed themselves to anything more than this, so their energy and effort trails
off, as all of their other commitments (particularly, those they are funded to do as part of
their jobs) come back to the fore. Without a process to resolve this concern, the utility of
further multi-day WSSSPE workshops is unclear.


\section*{Acknowledgments} \label{sec:acks}

This material is based upon work supported by the National Science Foundation (ACI-1648293, ACI-1547611),
by the Gordon and Betty Moore Foundation (GBMF\#5620),
and by the Alfred P.~Sloan Foundation (G-2016-7214).


\newpage
\appendix
\section{Organizing committee}  \label{sec:orgcom}

The following is the list of WSSSPE4 organizers.

{\scriptsize
\begin{longtable}{lll}
Gabrielle Allen &  University of Illinois Urbana--Champaign, USA \\
Jeffrey Carver  & University of Alabama, USA \\
Sou-Cheng (Terrya) Choi &  Illinois Institute of Technology, USA \\
Tom Crick & Cardiff Metropolitan University, UK \\
Michael R.~Crusoe & Common Workflow Language Project \\
Sandra Gesing & University of Notre Dame, USA \\
Robert Haines & University of Manchester, UK \\
Michael Heroux & Sandia National Laboratories, USA \\
Lorraine J.~Hwang & University of California, Davis, USA \\
Daniel S.~Katz & University of Illinois Urbana--Champaign, USA \\
Kyle E.~Niemeyer & Oregon State University, USA \\
Manish Parashar & Rutgers University, USA \\
Colin C.~Venters & University of Huddersfield, UK

\end{longtable}
}

\section{Attendees}  \label{sec:attendees}
The following is a list of attendees at WSSSPE4.

{\scriptsize
\begin{longtable}{lll}
Mark Abraham & SciLifeLab, KTH, Sweden\\
Aseel Aldabjan & University of Manchester\\
Alice Allen & Astrophysics Source Code Library\\
Gabrielle Allen & NCSA/University of Illinois\\
Pau Andrio & Barcelona Supercomputing Center\\
Wolfgang Bangerth & Colorado State University\\
Radovan Bast & UiT The Arctic University of Norway, Nordic e-Infrastructure Collaboration\\
Roman Bauer & Newcastle University\\
Steven Brandt & Louisiana State University\\
Lukas Johannes Breitwieser & CERN\\
Emily Chen & NCSA/University of Illinois Urbana--Champaign\\
Bruce Childers & University of Pittsburgh\\
Mike Croucher & University of Sheffield\\
Michael R. Crusoe & Common Workflow Language Project\\
Stephan Druskat & Humboldt-Universit\"{a}t zu Berlin\\
Anshu Dubey & Argonne National Laboratory\\
Tim Dunne & KnowInnovation\\
Peter Elmer & Princeton University\\
Iain Emsley & University of Oxford\\
Hans Fangohr & University of Southampton\\
Stuart Geiger & Berkeley Institute for Data Science\\
Sandra Gesing & University of Notre Dame\\
Carole Goble & University of Manchester, Software Sustainability Institute\\
Tom Goodale & Independent\\
Christopher Gwilliams & Cardiff University\\
Willem van Hage & Netherlands eScience Center\\
Robert Haines & University of Manchester\\
James Hetherington & University College London\\
Simon Hettrick & Software Sustainability Institute\\
Neil Chue Hong & Software Sustainability Institute\\
Lorraine Hwang & UC Davis, CIG\\
Ray Idaszak & RENCI, University of North Carolina at Chapel Hill\\
Caroline Jay & University of Manchester\\
Brian Jimenez & Barcelona Supercomputing Center\\
Rafael C Jimenez & ELIXIR Hub, UK\\
Daniel S Katz & University of Illinois Urbana--Champaign\\
Dominic Kempf & Heidelberg University\\
Frank L\"{o}ffler & Louisiana State University\\
Patricia Lago & Vrije Universiteit Amsterdam\\
Xinlian Liu & Hood College\\
Hannes Loeffler & STFC\\
Tengyu Ma & Vanderbilt University\\
Robert H McDonald & Indiana University\\
Chris Mentzel & Moore Foundation\\
Constantinos Michailidis & KnowInnovation\\
Jonah Miller & Perimeter Institute of Theoretical Physics\\
Daniel Mosse & University of Pittsburgh\\
August (Gus) Muench & American Astronomical Society\\
Santiago N\'u\~nez-Corrales & University of Illinois Urbana--Champaign\\
Udit Nangia & University of Illinois Urbana--Champaign\\
Aleksandra Nenadic & University of Manchester, SSI\\
Kyle Niemeyer & Oregon State University\\
Dominic Orchard & University of Kent\\
Francisco Queiroz & Pontifical Catholic University of Rio de Janeiro\\
Fons Rademakers & CERN\\
Karthik Ram & University of California, Berkeley\\
Sam Relton & University Of Manchester\\
Tom Robitaille & Freelance\\
Jean Salac & University of Virginia\\
Janos Sallai & Vanderbilt University\\
Prabhjyot (Singh) Saluja & Southwestern Oklahoma State University\\
Eric Seidel & University of California, San Diego\\
Sameer Shende & University of Oregon\\
Raniere Silva & University of Manchester\\
Shoaib Sufi & University of Manchester, Software Sustainability Institute\\
Chris Sweet & University of Notre Dame\\
Rob Welch & University of Leeds\\
Christopher Woods & University of Bristol\\

\end{longtable}
}

\section{Travel award recipients}  \label{sec:awardees}
The following table contains the list of WSSSPE4 travel award recipients, where
\textsuperscript{*} and \textsuperscript{\textdagger} indicate students and
early-career researchers, respectively.

{\scriptsize
\begin{longtable}{lll}
Alice Allen & Astrophysics Source Code Library \\
Wolfgang Bangerth & Colorado State University, USA \\
Roman Bauer\textsuperscript{\textdagger} & Newcastle University, UK \\
Steven Brandt & Louisiana State University, USA \\
Emily Chen\textsuperscript{*} & University of Illinois Urbana-Champaign, USA \\
Bruce Childers & University of Pittsburgh, USA \\
Michael R.~Crusoe\textsuperscript{\textdagger} & Common Workflow Language Project \\
Stephan Druskat\textsuperscript{\textdagger} & Humboldt-Universit\"{a}t zu Berlin, Germany \\
Anshu Dubey & Argonne National Laboratory, USA \\
Iain Emsley\textsuperscript{*} & Oxford University, UK \\
Christopher Gwilliams\textsuperscript{\textdagger} & Cardiff University, UK \\
Lorraine Hwang & University of California Davis, USA \\
Ray Idaszak & RENCI, University of North Carolina at Chapel Hill, USA \\
Xinlian Liu & Hood College, USA \\
Frank Löffler & Louisiana State University, USA \\
Tengyu Ma\textsuperscript{*} & Vanderbilt University, USA \\
Robert H.~McDonald & Indiana University, USA \\
Jonah Miller\textsuperscript{*} & Perimeter Institute, Canada \\
Daniel Mosse & University of Pittsburgh, USA \\
Kyle E.~Niemeyer\textsuperscript{\textdagger} & Oregon State University, USA \\
Santiago N\'{u}\~{n}ez-Corrales\textsuperscript{*} & University of Illinois Urbana-Champaign, USA \\
Dominic Orchard & University of Kent, UK \\
Francisco Queiroz\textsuperscript{*} & Pontifical Catholic University of Rio de Janeiro, Brazil \\
Jean Salac\textsuperscript{*} & University of Virginia, USA \\
Janos Sallai & Vanderbilt University, USA \\
Prabhjyot (Singh) Saluja\textsuperscript{*} & Southwest Oklahoma State University, USA \\
Eric Seidel\textsuperscript{*} & University of California San Diego, USA \\
Karan Shah\textsuperscript{*} & Georgia Institute of Technology, USA \\
Sameer Shende & University of Oregon, USA

\end{longtable}
}

\section{Program committee}  \label{sec:progcom}
The following table lists the WSSSPE4 program committee.

{\scriptsize
\begin{longtable}{lll}
David Abramson & University of Queensland, Australia\\
Lorena A. Barba & George Washington University, USA\\
Ross Bartlett & Sandia National Laboratories, USA\\
Christoph Becker & University of Toronto, Canada\\
Ewout van den Berg & IBM Watson, USA\\
David Bernholdt & Oak Ridge National Laboratory, USA\\
Stefanie Betz & Karlsruhe Institute of Technology, Germany\\
Coral Calero & Universidad Castilla La Mancha, Spain\\
Ishwar Chandramouli & National Cancer Institute, National Institutes of Health, USA\\
Ruzanna Chitchyan & University of Leicester, UK\\
Karen Cranston & Duke University, USA\\
Ewa Deelman & Information Sciences Institute, University of Southern California, USA\\
Charlie E. Dibsdale & O-Sys, Rolls Royce PLC, UK\\
Anshu Dubey & Argonne National Laboratory, USA\\
Nadia Eghbal & Independent Researcher (via Ford Foundation), USA\\
Peter Elmer & CERN, Switzerland\\
Martin Fenner & DataCite, Germany\\
David Gavaghan & University of Oxford, UK\\
Mike Glass & Sandia National Laboratories, USA\\
Carole Goble & University of Manchester, UK\\
Joshua Greenberg & Alfred P. Sloan Foundation, USA\\
Michael K Griffiths & University of Sheffield, UK\\
Sarah Harris & University of Leeds, UK\\
James Hetherington & University College London, UK\\
Fred J. Hickernell & Illinois Institute of Technology, USA\\
Neil Chue Hong & Software Sustainability Institute, University of Edinburgh, UK\\
Caroline Jay & University of Manchester, UK\\
Rafael C. Jimenez & ELIXER, Cambridge, UK\\
Matthew B. Jones & University of California Santa Barbara, USA\\
Nick Jones & New Zealand eScience Infrastructure (NeSI), NZ\\
Sedef Akinli Kocak & Ryerson University, Canada\\
Jong-Suk Ruth Lee & National Institute of Supercomputing and Networking, KISTI, Korea\\
James Lin & Shanghai Jiao Tong University, China\\
Frank Löffler & Louisiana State University, USA\\
Gregory Madey & University of Notre Dame, USA\\
Ketan Maheshwari & University of Pittsburgh, USA\\
Steven Manos & University of Melbourne, Australia\\
Chris A. Mattmann & NASA JPL; University of Southern California, USA\\
Abigail Cabunoc Mayes & Mozilla Science Lab, USA\\
Robert H. McDonald & Indiana University, USA\\
Lois Curfman McInnes & Argonne National Laboratory, USA\\
Alberto Di Meglio & CERN, Switzerland\\
Chris Mentzel & Gordon and Betty Moore Foundation, USA\\
Peter Murray-Rust & University of Cambridge, UK\\
Christopher R. Myers & Cornell University, USA\\
Jarek Nabrzyski & University of Notre Dame, USA\\
Cameron Neylon & Curtin University, Australia\\
Aleksandra Pawlik & New Zealand eScience Infrastructure (NeSI), NZ\\
Fernando Perez & Lawrence Berkeley National Laboratory; University of California, Berkeley, USA\\
Marian Petre & The Open University, UK\\
Marlon Pierce & Indiana University, USA\\
Andreas Prlic & University of California, San Diego, USA\\
Karthik Ram & University of California, Berkeley, USA\\
Morris Riedel & Juelich Supercomputing Centre, Germany\\
Dave De Roure & Oxford e-Research Centre, University of Oxford, UK\\
Norbert Seyff & University of Zurich, Switzerland\\
Arfon Smith & GitHub Inc, USA\\
Borja Sotomayor & University of Chicago, USA\\
Edgar Spalding & University of Wisconsin, USA\\
Maria Spichkova & RMIT University, Australia\\
Victoria Stodden & University of Illinois Urbana--Champaign, USA\\
Matthew Turk & University of Illinois Urbana--Champaign, USA\\
Nancy Wilkins-Diehr & San Diego Supercomputer Center, University of California, San Diego, USA\\
James Willenbring & Sandia National Laboratories, USA\\
Scott Wilson & Cetis LLP, UK\\
Theresa Windus & Iowa State University and Ames Laboratory, USA\\

\end{longtable}
}

\section{Detailed survey results}  \label{sec:survey_details}

Note: typos and spelling errors have been corrected, and spelling has been changed from UK standard to US standard in some cases, but no other changes have been made to the attendee responses.

\begin{enumerate}
\item \textbf{Please indicate how strongly you agree or disagree with the following statements}

\begin{table}[h!]
\centering
\caption{General questions about the conference program}
\label{tab:survey_program}
{\scriptsize
\begin{tabular}{@{}lcccccc@{}}
\toprule
\multirow{2}{*}{Question} & strongly & somewhat & agree & somewhat & strongly &
\multirow{2}{*}{Total} \\
& disagree & disagree & & agree & agree & \\
\midrule
\rowcolor{lightgray} The plenary sessions had a good & & & & & & \\
\rowcolor{lightgray} mix of topics and speakers. &
\multirow{-2}{*}{5} &
\multirow{-2}{*}{3} &
\multirow{-2}{*}{8} &
\multirow{-2}{*}{11} &
\multirow{-2}{*}{18} &
\multirow{-2}{*}{45} \\
The balance was right between &
\multirow{2}{*}{6} &
\multirow{2}{*}{2} &
\multirow{2}{*}{7} &
\multirow{2}{*}{9} &
\multirow{2}{*}{21} &
\multirow{2}{*}{45} \\
talks, panels, and lightning talks. & & & & & & \\
\rowcolor{lightgray} There was sufficient time for discussion. &
6 &
6 &
8 &
9 &
16 &
45 \\
\bottomrule
\end{tabular}
}
\end{table}


\begin{figure}[!htb]
    \captionsetup[subfigure]{position=b}
    \centering
    \subcaptionbox{
    The plenary sessions had a good mix of topics and speakers.
    \label{fig:SFig1}
    }{\includegraphics[width=.33\linewidth]{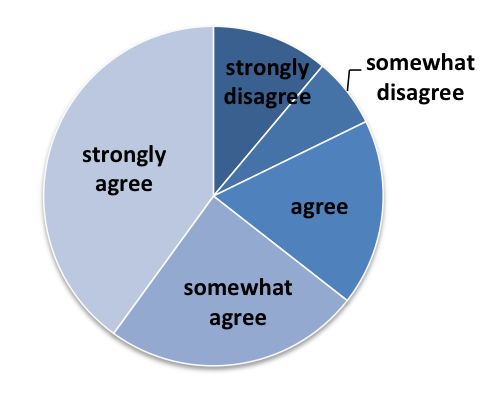}}
    ~
    \subcaptionbox{
    The balance was right between talks, panels, and lightning talks.
    \label{fig:SFig2}
    }{\includegraphics[width=.32\linewidth]{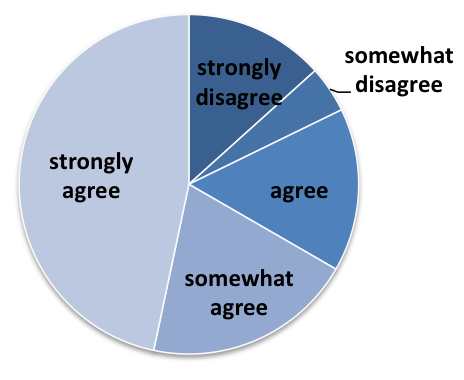}}
    ~
    \subcaptionbox{
    There was sufficient time for discussion.
    \label{fig:SFig3}
    }{\includegraphics[width=.26\linewidth]{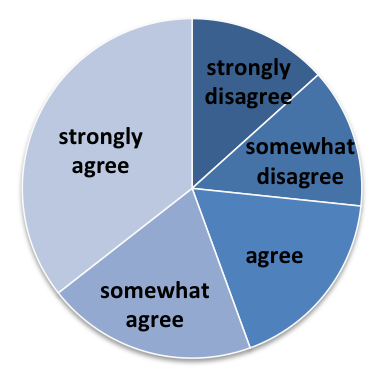}}
    \caption{Responses to general questions about the conference program}
\end{figure}

\item \textbf{What part of the meeting was the most valuable? select all that apply}

\begin{figure}[h!]
    \centering
    \begin{tabular}{@{}c@{}}
        \includegraphics[width=0.35\textwidth]{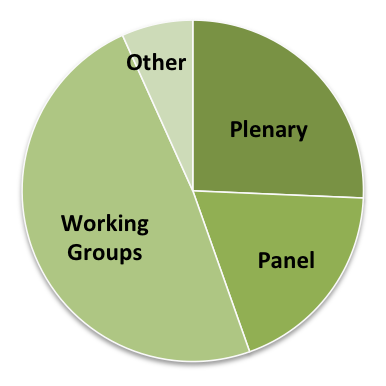}
    \end{tabular}
    \qquad
    \rowcolors{1}{}{lightgray}
    \begin{tabular}{@{}lc@{}}
    \toprule
    Answer & Count \\
    \midrule
    Plenary & 19 \\
    Panel & 14 \\
    Working Groups & 36 \\
    Other & 5 \\
    Total & 74 \\
    \bottomrule
    \end{tabular}
    \caption{Responses to ``What part of the meeting was the most valuable?''}
    \label{fig:SFig4}
\end{figure}

\item \textbf{What additional topics would you like to see discussed at the next meeting?}

\begin{itemize}
\item Concrete reports about best practices
\item considerations of software licensing, build systems, delivery (source vs binary vs whatever) - but of course these are ``later'' issues... first one needs the software to not be hidden, to have version control, have a README, etc.
\item Container sustainability practices
\item education, training interns
\item Explicit follow-up at WSSSPE5 on all the activities that were created at WSSPE4.  These should be at least lightning talks, perhaps with their own session.
\item financial models for sustaining software
\item Frameworks for building software
\item have a short beginning presentation for each working group and a short ending presentation
\item How software development training can be adapted to different disciplines beyond STEM
\item How to low the barrier for non-tech heavy researchers to follow best software practices.
\item I would like to have more discussions about research, if people can share their research or some of the interesting projects that they're working on, this will help in knowing the problems in sustaining software.
\item I would like to learn more about sustainability issues in industry.
\item I would like to see more grounded advice and less blue-sky thinking. A lot of stuff in the talks was `this would be nice' but there were not enough talks that were directly applicable to day-to-day programming.
\item Like to see more examples from the real world.
\item Mechanisms of improving software reusability
\item More concrete projects, concrete possibilities and showcases for collaboration with results
\item Practical ways of achieving some of the goals rather than the discussion of more attempts on the writing of white papers
\item Reproducibility through containerization (I guess there is another list of best practices needed there)
\item RSE award, organizational questions regarding setting up RSE teams within universities
\item Software testing
\item Some theory on what makes for sustainable software, which is not the same as best practice for research software engineering. If this had been talked about before then consider that 80\% of us were new.
\item Specific scientific software sustainability factors
\item Testing methods that work. Ways of promoting organizational change.
\item The progress of our current project.
\item The use of scientific software at different stages of research / work
\item We discussed scientific software. Should we also discuss software FOR scientists (like Dan's google scholar replacement)?
\end{itemize}

\item \textbf{What could be improved?}

\begin{itemize}
\item A bit less fly by night scheduling though things worked out just fine.
\item A lot of the speakers just seemed to be there either to plug their own projects (e.g. Hydroshare, BioDynaMo) or spent the majority of their talk waffling (e.g. Manifesto for Personal Responsibility in the Engineering of Academic Software). Also, the icebreakers were not very useful.
\item A single room for the duration of the meeting. (Particular to this WSSSPE4, the ``collaboratory'' room with the monitors, was actually perfect for the full meeting.)
\item Allotting more time for questions \& discussions in between talks
\item close to perfect workshop - if I would change one thing that give more time for discussions
\item consistency of participants
\item felt locked in to the breakout group, not sure if that was intended or not
\item Finding rooms a bit confusing at the meeting outset
\item Having more time for the demos.
\item Having more time to discuss some of the talks
\item I was disappointed in the fruit plate because it only had melon.
\item I would like to be able to ask questions after talks. By the time we can discuss the talks, I've forgotten what I wanted to say
\item Identify a local pub/restaurant/cafe to which people can head afterwards for casual meetup for dinner/drinks/discussion/whatever
\item If feedback to people's projects and lighting talks etc is posted online, then we would have also benefits from those thoughts
\item It was a great event overall I feel like if we can bring some real life project integration with best practices that will be helpful
\item it would be great to see some specific case studies/success stories
\item Keep the online agenda up to date. Better air-conditioning.
\item Less talks
\item more time for discussion and working groups
\item not sure, it looks great to me. I would like to see in plenary the perspective from closed software and open source software (for cross-fertilization)
\item Some longer talks that do deeper dives.  Good that the parallel track idea was abandoned so we could hear all talks.
\item The identification of what RSE and RS are, especially for newcomers
\item The keynote address could have been better targeted for this community beforehand.
\item The workshop was great!!!
\item There is lots of vision here, so we need to put more effort in our project to move forward.
\item Think about a little more planning on the BOF Dinners to make it easier to get groups together. Maybe a Google doc to encourage it etc.
\item To have a few speakers (2-3) that talk about concrete scientific software, showing the simulations and talking about their scientific results, and why they chose the software they did (as compared to other available ones)
\item Try and add some extra Q\&A time after talks (unless that'd mean less talks)
\item We could have additional time to work in working groups
\item WSSPE vision prior to start would have helped
\end{itemize}

\item \textbf{What would make you more likely to invest time in future WSSSPE initiatives?}

\begin{itemize}
\item A centralized place/organization to see and discuss WSSSPE initiatives. At the moment, WSSSPE organization is scattered across many places. One website with a list of activities and contributors would help.
\item A tutorial session.
\item Being able to contribute to something going forward, rather than a talking shop for the initiatives.
\item connect to people
\item For meetings: again the availability of travel awards.
\item For myself, I anticipate looking for ways to tie WSSPE initiatives with my current projects such that the outcome benefits a/the broader community.
\item Free lunch and double coffee capacity? Also - seriously - less filler talks and filler activities.
\item Funded support resources
\item funding; focus on more concrete aspects of sustainability
\item Having more time to invest in future WSSSPE initiatives
\item Having more time to invest in future WSSSPE initiatives; I'd definitely do more if I could.
\item Honestly, having the time available in my own schedule (which I don't think is something WSSSPE organizers can change)
\item I already plan to do it :)
\item I'm not sure. I'm not very interested in fostering WSSSPE as an thing, but I am interested in WSSSPE as a means to network to find people to work on these issues with me.
\item If I can see the value of the initiative to me and my value to the initiative
\item if I had funding to specifically work on that
\item If my daily workload would be reduced
\item if the community is and remains active
\item Keeping up your generous travel support alone is enough for me to not even think about missing any further WSSSPE events
\item Know that the work started in one edition can be presented on the following one.
\item Likelihood of useful new contributors.
\item more clearly defined groups
\item More followups between the conferences.
\item More funding opportunities to be able to devote appropriate attention to the topic
\item More senior participants with (i) experience, (ii) influence
\item More time. Some funding.
\item Potential a
\item Prospects for collaborations leading to citable artifacts, e.g. papers
\item some documented forward momentum between sessions
\item The networking is incredible for this meeting and should be encouraged to continue. Having published deliverables out of the community will help this as well.
\item The travel support really helps
\item This topic is related to a large component of what I envision as my professional career path. In particular, the project designed during the sessions will be fleshed out into a demo, and if adequate, into some form of larger piece of software.
\item To establish a metrics for research software.
\item Understanding the landscape of funding and potential for publication
\item Yes, definitely I met with some very interesting people through this platform and would like to be a part of it. It'll be an honor to collaborate to this initiative.
\end{itemize}

\item \textbf{Would you be interested in joining a professional society focused around the topics of WSSSPE?}

\begin{figure}[h!]
    \centering
    \begin{tabular}{@{}c@{}}
        \includegraphics[width=0.35\textwidth]{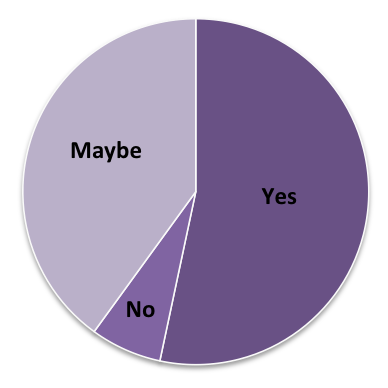}
    \end{tabular}
    \qquad
    \rowcolors{1}{}{lightgray}
    \begin{tabular}{@{}lc@{}}
    \toprule
    Answer & Count \\
    \midrule
    Yes & 24 \\
    No & 3 \\
    Maybe & 18 \\
    Total & 45 \\
    \bottomrule
    \end{tabular}
    \caption{Responses to ``Would you be interested in joining a professional
    society focused around the topics of WSSSPE?''}
    \label{fig:SFig5}
\end{figure}

%

\item \textbf{Will you be part of a working group with colleagues going forward on a topic addressed at WSSSPE, and if so, which one?}

\begin{figure}[h!]
    \centering
    \begin{tabular}{@{}c@{}}
        \includegraphics[width=0.35\textwidth]{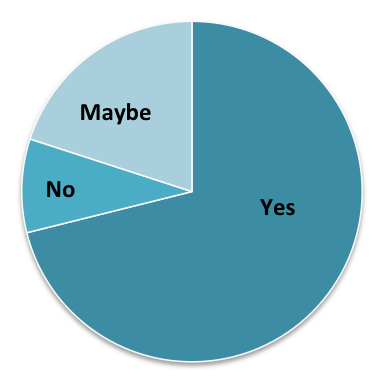}
    \end{tabular}
    \qquad
    \rowcolors{1}{}{lightgray}
    \begin{tabular}{@{}lc@{}}
    \toprule
    Answer & Count \\
    \midrule
    Yes & 32 \\
    No & 4 \\
    Maybe & 9 \\
    Total & 45 \\
    \bottomrule
    \end{tabular}
    \caption{Responses to ``Will you be part of a working group  with colleagues
    going forward  on a topic addressed at WSSSPE?''}
    \label{fig:SFig6}
\end{figure}



\FloatBarrier
\item \textbf{Working group name}


\begin{itemize}
\item ``Citations / CodeMeta''  and  ``Software Practices White Paper''
\item Best Practices White Paper and Best Practices Undergraduate Course
\item CodeMeta
\item CodeMeta
\item CodeMeta
\item CodeMeta, but I also supplied information to other working groups -- mostly connections to others -- that I hope will be of assistance.
\item Continuous CI for research software on uncommon hardware
\item Continuous Integration
\item Continuous integration for research software; Testing of scientific software
\item Curriculum for an undergraduate course on software best practices for domain scientists
\item Domain Specific Undergraduate Courses on Best Practices for Software Development
\item Domain Specific Undergraduate Courses on Best Practices for Software Development AND International Collaborations between institutions.
\item integration of software testing through Debian
\item Meaningful Metrics for Scientific Software
\item Meaningful Metrics in Software
\item metadata
\item multiple
\item Open Research Index, Metrics, Social Science, and maybe others
\item Prototypes
\item research software studies
\item Scientific Software Prototyping Infrastructure (S2PI), WG-Best-Practices
\item Setting up Software Sustainability Alliance
\item Software Sustainability Alliance
\item Software Sustainability Alliance; Metrics; White Paper
\item sustainability metrics; best practices
\item Sustainable Software Alliance
\item Sustainable Software Alliance
\item Tenure Letters
\item Testing in Sciences
\item Testing in Scientific Software
\item Verify Best Practices and Metrics for Sustainable Software
\item Verify Best Practices Case Studies
\item White Paper WG, Open Research Index WG
\end{itemize}

\item \textbf{Do you have any other comments or suggestions or ideas?}

\begin{itemize}
\item As a student, coming to WSSSPE4 has been very rewarding experience. Sustainable software is not my particularly sub-field of interest, but it was very interesting to me to apply the skills I had learned in my research in a different field. It definitely taught me a different perspective and expanded my original perception of where my skills can be applied.
\item As a student, this has helped me explore the different computational research careers and has made me more aware of scientific software best practices.  I found the plenary talks particularly valuable for learning about different areas of study, and it was great how willing all of the WSSSPE participants were to share more and help each other. I also found it valuable that there were other undergraduates at this WSSSPE, and it created another small community and encouraged me to participate more.
\item buddy me-sheet and workshop planning and moderation was excellent
\item Don't start at 8.45am if you're asking people to socialize and connect in the evenings! Work on getting a better gender, geographical and discipline balance. The use of travel bursaries has helped (particularly in getting early career participants).
\item Great organization, enjoyed being here very much, it was surely a success!
\item Great to meet colleagues from the UK and Germany at this event and enjoyed this format for working, engaging, and producing with the community while at the workshop that was facilitated by KnowInnovation.
\item I enjoyed the brainstorming session on Tuesday morning - well organized
\item I had hoped to have some discussion of building the sustainable communities beyond the software writing stage. The HPC focus, though expected slightly, was too heavy and I think ignores some of the other long term issues, such as preservation and addressing first principles for new technologies, or smaller, individual projects that will also support science. Perhaps it would be good to have a student session as well? (But I do like the integration).
\item I think the mixing of students, staff, faculty etc in the working groups is very effective and will contribute to  outcomes outside of the working group. It was great to see that we have undergraduates leading working groups here. There should be more attention on diversity.
\item I wish there were a project on training interns to work at academic environment as programmers
\item I would like to thank the committee for selecting me. As an undergraduate student this was a great experience I got to meet professors from the universities which I wanted to join. Now I exactly know how to prepare my application. I learned some really good technical skills and will definitely apply it to my research. So happy to be a part of this wonderful group. I will design a  logo for WSSSPE next year. Hope I can attend this event next year also.
\item Is there something better than sharpies (few of running around the room with a permanent marker)? Travel Awardee: Discussions here will and have benefitted my organization and community.
\item It was great to connect with people doing research software engineering in various positions, contexts, and universities. I felt like I learned a lot. Moving between the two rooms was a bit confusing.
\item Maybe involve more researchers from low-income countries if funds are available.
\item no. I think the moderator and the format are both great.
\item Nothing comes to mind
\item Receiving a travel award was crucial for me to come to WSSSPE4 this time, especially because WSSSPE4 was held outside the US. On the other hand, I see the great benefit this had on the geographic diversity of the participants, so I do think that meetings not only in the US are of great value. And again, I would like to emphasize, that travel awards are crucial for this type of meeting, as most participants are not directly supported for this type of activity - one reason to have a WSSSP5. (on another note: have a multi-line text field for comments next time)
\item The issue of software sustainability is on par with the challenges of building scientific instruments. Part of what I am for my doctoral dissertation is finding better abstractions, which leads to sustainability through better mathematics. This workshop raised my awareness in several other aspects, which will be considered for this and all future projects I may be related to.
\item The travel award made it possible for me to come to WSSSPE-4 and broaden my horizons.  Many issues pertaining research software and sustainable software were uncovered, and WSSSPE-4 brought to light the importance of the issues.  I'd encourage other travel awards to people, like me, who do not belong to the community.
\item There aren't lots of researchers knowing such kind of concept, sustainable software. We need to let more RSE know it.
\item There is a lot of discussion, but it's hard to see what actually will come of it. The discussion of the 9 working groups from last year at the beginning of the workshop clearly showed this: IIRC, not one had produced anything, all seemed to be stalled somewhere. We ought to find ways to actually make things *happen* after we talk about it (and admittedly have many good ideas).
\item This has been my first WSSSPE, and as someone coming from the humanities with an interest in RSE it has been an enormously rewarding experience to be able to interact with and learn from individuals and initiatives that have been part of the greater discussion around WSSSPE topics more in both quantity and quality. For me personally, it made me aware that I really need to refactor a lot of my basic assumptions for a PhD thesis, but that's a great thing because it means that I was able to draw on the exchanges and experiences from WSSSPE already. All in all, the workshop has made me more determined to become more involved with this community. And last but not least i just wanna say THANK YOU for the generous travel support I've received, which meant that my PI and dept. didn't have to have any doubts whether I'd drain their resources which are meant to be used towards achieving a completely different research goal.
\item This was an excellent mix of people, timed/structured activities, and untimed events.
\item This was an excellent mix of people, timed/structured activities, and untimed events. I had the opportunity to talk to numerous people about projects, issues, and ways to improve science/scientific software, and leave with many new ideas and possible collaborations/cooperative efforts with complementary projects. The working groups activities were excellent, a good mix of projects, and I saw a lot of progress on several I'm very interested in. (Please group this survey with my earlier one; I hit ``end'' too soon on the other! Thanks!)
\item We have activity on slack, github, and twitter. And working groups have their own stuff. Can we index these in a single place?
\item What percentage of the working groups actually end up producing anything? It seems to me as though most come up with an idea and then find out they don't have the resources to implement it - nice as the ideas are. Also, I assume someone is going to attempt to reinvent metrics or best practices every single year, I guess that's unavoidable.
\item WSSPE seems to be evolving nicely from year to year.  I hope that in WSSSPE5 there are lots of results to report from the outcome of WSSSPE4 working groups.
\item You could try unconferences instead of tracks.
\end{itemize}

\end{enumerate}

\newpage
\bibliographystyle{vancouver}

\bibliography{wssspe}
\end{document}